\documentclass[aos,preprint]{imsart}

\RequirePackage{amsthm,amsmath,amsfonts,amssymb}
\RequirePackage[numbers,sort&compress]{natbib}
\RequirePackage[colorlinks,citecolor=blue,urlcolor=blue]{hyperref}
\RequirePackage{graphicx}

\usepackage{bm}
\usepackage{natbib}
\usepackage{xcolor}
\usepackage{amsfonts}
\usepackage{amsmath}
\usepackage{amssymb}
\usepackage{mathrsfs}
\usepackage{hyperref}
\usepackage{booktabs}
\usepackage{graphicx}
\usepackage{multirow, multicol}
\usepackage{algorithm}
\usepackage{algpseudocode}
\usepackage{etoolbox}
\AtBeginEnvironment{algorithmic}{\setcounter{ALG@line}{0}}
\usepackage{enumitem}
\usepackage{epstopdf}
\startlocaldefs
\theoremstyle{plain}

\newtheorem{theorem}{Theorem}

\newtheorem{corollary}{Corollary}
\newtheorem{prop}{Proposition}
\newtheorem{assumption}{Assumption}
\theoremstyle{remark}
\newtheorem{definition}[theorem]{Definition}
\newtheorem{example}{Example}

\newcommand{\blamb}{\boldsymbol{\lambda}}
\newcommand{\bit}{\begin{itemize}}
	\newcommand{\eit}{\end{itemize}}
\newcommand{\ba}{\begin{eqnarray}}
	\newcommand{\ea}{\end{eqnarray}}
\newcommand{\bas}{\begin{eqnarray*}}
	\newcommand{\eas}{\end{eqnarray*}}
\newcommand{\ben}{\begin{enumerate}}
	\newcommand{\een}{\end{enumerate}}
\newcommand{\e}{ { \mathbb{E}}}

\newcommand{\cov}{ {\mathbb{C}\rm ov}}

\newcommand{\bx}{{\bf x}}
\newcommand{\by}{{\bf y}}
\newcommand{\bX}{{\bf X}}

\newcommand{\bZ}{{\bf Z}}

\newcommand{\bW}{{\bf W}}
\newcommand{\bw}{{\bf w}}
\newcommand{\bz}{{\bf z}}
\newcommand{\bY}{{\bf Y}}
\newcommand{\bu}{{\bf u}}
\newcommand{\bmu}{{\bm\mu}}
\newcommand{\bv}{{\bf v}}
\newcommand{\bU}{{\bf U}}
\newcommand{\bO}{{\bf O}}

\newcommand{\bp}{{\bm p}}
\newcommand{\bg}{{\bm g}}

\newcommand{\bI}{{\bm I}}
\newcommand{\bJ}{{\bm J}}
\newcommand{\bt}{{\bm t}}

\newcommand{\bq}{{\bm q}}

\def\T{{ \mathrm{\scriptscriptstyle \top} }}

\newcounter{aqctr}
\newenvironment{author-query}
{\refstepcounter{aqctr}\par\vspace{\baselineskip}\noindent
\color{red}\textbf{Author Query/Comment AQ \arabic{aqctr}.}}
{\par\vspace{\baselineskip}\normalcolor}

\endlocaldefs

\begin{document}

\begin{frontmatter}
\title{Deep conditional distribution estimation via nonparametric exponential tilt models}
\runtitle{Deep conditional distribution estimation}

\begin{aug}
\author[A]{\fnms{Junjun}~\snm{Lang}{}},
\author[B]{\fnms{Yukun}~\snm{Liu}\thanksref{t1}},
\and
\author[C]{\fnms{Jing}~\snm{Qin}{}}
\thankstext{t1}{Corresponding author. Email: \href{mailto:ykliu@sfs.ecnu.edu.cn}{ykliu@sfs.ecnu.edu.cn}.}
\address[A]{School of Statistics and Data Science, Nankai University, Tianjin 300071, China}
\address[B]{KLATASDS-MOE, School of Statistics,
 East China Normal University,
 Shanghai 200062, China}
\address[C]{National Institute of Allergy and Infectious Diseases, National Institutes of Health, Bethesda, MD 20892, USA}
\end{aug}

\begin{abstract}
Estimating the conditional distribution of a response $\bY$ given covariates $\bX$ is a fundamental problem in statistics, underlying prediction, feature estimation, and dependence assessment. We propose a unified nonparametric representation of the conditional density, a nonparametric exponential tilt model, in which the dependence between $\bX$ and $\bY$ is captured by an interaction function $\phi(\bx,\by)$ while both baseline densities are left unspecified. We establish identifiability and develop a nonparametric likelihood method that profiles out the baselines by empirical likelihood, with non-asymptotic error bounds under general sieve conditions.
Because the profile likelihood  admits no closed-form or readily tractable population counterpart, standard empirical process arguments do not apply; we therefore establish a new exponential tail bound for the supremum of a canonical U-process under a uniform entropy condition, a result we show to be new to the U-process literature and of independent interest, with a largely self-contained proof.
The framework unifies several inferential tasks: mutual density ratio and mutual information estimation; a general theory for conditional features, including means, quantiles and distribution functions, without separate smoothness assumptions on the individual conditional feature functions; and a finite-sample valid permutation test of independence. Under deep neural network sieves, the convergence rates depend on the intrinsic complexity of $\phi$ rather than the ambient dimension. Simulations demonstrate that the proposed method attains uniformly smaller estimation errors than existing  conditional density and mutual density-ratio estimators, and delivers markedly higher power in independence testing when dependence is oscillatory or high-dimensional, while remaining comparable to established tests in standard settings.
\end{abstract}



\end{frontmatter}

\section{Introduction}
\label{sec-intro}

\subsection{Background}

Estimating the conditional distribution of a response $\bY$ given a covariate vector $\bX$ lies at the heart of statistical inference: it underpins prediction, classification, and the principled quantification of dependence between random variables. The classical approach is parametric, and the most widely used parametric framework is the generalized linear model (GLM) of \cite{nelder1972glm}; see \cite{mccullagh1989glm} for a comprehensive treatment. For a scalar response $Y$, a GLM assumes that the conditional density of $Y$ given $\bX=\bx$ belongs to a one-parameter exponential family,
\begin{equation}\label{eq:glm}
	f(y\mid \bx)=f_*(y)\exp\{y\,\theta(\bx)-\psi(\theta(\bx))\},
\end{equation}
where $f_*$ is a specified baseline (carrier) density, $\theta(\bx)$ is the natural parameter, and $\psi(\theta)=\log\int e^{y\theta}f_*(y)\,dy$ is the cumulant function that normalizes the density. The natural parameter is tied to the covariates through a linear predictor, $\theta(\bx)=\eta(\bx^{\top}\beta)$, for a known function $\eta$. When $\eta$ is the identity, so that $\theta(\bx)=\bx^{\top}\beta$, the model uses the canonical link: the conditional mean $\mu(\bx)=\psi'(\theta(\bx))$ then satisfies $g(\mu)=\bx^{\top}\beta$ with $g=(\psi')^{-1}$. Familiar special cases include the linear model with normal $f_*$ and identity link, logistic regression with Bernoulli $f_*$ and logit link, Poisson regression with log link, and Gamma regression with reciprocal link. Model \eqref{eq:glm} therefore imposes two parametric restrictions: a known baseline $f_*$ and a linear (or index) form for the natural parameter.

Much subsequent work has relaxed one of these two restrictions while retaining the other. The first direction keeps the exponential-family baseline but replaces the linear predictor with a flexible function of the covariates; examples include generalized additive models \citep{hastie1990gam} and single-index models \citep{ichimura1993}. More recently, deep neural networks (DNNs) have been employed for this purpose, building on the theory for nonparametric regression with DNN sieves \citep{bauer2019, schmidtHieber2020}. In particular, \cite{meng2026} replaced $\eta(\bx^{\top}\beta)$ in \eqref{eq:glm} by a DNN and, through simulations under nonparametric logistic, Poisson, and binomial regression models, demonstrated the effectiveness and efficiency of this approach.

The second direction keeps the index structure $\theta(\bx)=\bx^{\top}\beta$ but leaves the baseline $f_*$ unspecified. This is the idea behind the density ratio (or exponential tilt) model \citep{anderson1979, qin1997, qin1998, diao2012} and the proportional likelihood ratio model of \cite{luo2012}. Because the baseline is an infinite-dimensional nuisance parameter, it is profiled out by empirical likelihood \citep{owen1988, owen2001, qin1994} or by nonparametric profile likelihood, yielding semiparametric estimators of $\beta$ that are robust to misspecification of $f_*$.

Nevertheless, the extensions discussed above share two notable limitations. First, they are developed primarily for a scalar response $Y$ and may not directly apply to a multivariate response $\bY$. When the response is multivariate, the exponential-family specification in \eqref{eq:glm} becomes difficult to formulate or justify, and the density ratio and proportional likelihood ratio models require a multivariate baseline whose specification and profiling pose substantial challenges; moreover, the index structure $\bx^{\top}\beta$ is generally too restrictive to capture the dependence between a covariate vector and a multivariate response. Second, these approaches still impose structural assumptions on the conditional density or distribution of the response: the first direction retains a parametric baseline $f_*$, while the second retains a linear index structure. Whenever the true conditional distribution deviates from these assumptions, the resulting estimators can be biased and the associated inference unreliable. A misspecified model may yield distorted estimates of conditional densities, invalid standard errors and confidence intervals, and misleading conclusions about the relationship between $\bX$ and $\bY$, errors that can propagate into downstream scientific decisions. These considerations motivate a fully data-driven formulation that avoids such restrictions whenever possible and lets the data determine the form of the conditional distribution.

\subsection{Our contributions}

In this paper, we generalize model \eqref{eq:glm} along three directions simultaneously:
\ben
\item[(1)] the baseline density $f_*$ is left entirely unspecified, as in the density ratio and proportional likelihood ratio models;
\item[(2)] the bilinear term $y\,\theta(\bx)$ in the exponent is replaced by a general nonparametric interaction function $\phi(\bx,\by)$, so that no parametric or semi-parametric assumption is imposed on the covariate effect and the response is allowed to enter the exponent nonlinearly;
\item[(3)] the response $\bY$ is permitted to be multivariate rather than scalar, and $\bX$ and $\bY$ are treated symmetrically.
\een
The resulting conditional density of $\bY$ given $\bX$,
\ba\label{NETM}
f(\by\mid\bx)=\frac{f_2(\by)\exp\{\phi(\bx,\by)\}}{\int f_2(\bv)\exp\{\phi(\bx,\bv)\}\,d\bv},
\ea
defines a \emph{nonparametric exponential tilt model} (NETM): the unspecified baseline density $f_2$ is tilted by the factor $\exp\{\phi(\bx,\by)\}$, and the cumulant function $\psi$ in \eqref{eq:glm} is replaced by the normalizing integral in the denominator. It is easily verified that any conditional density \(f(\by\mid\bx)\) that is strictly positive on a common support admits the representation \eqref{NETM}. Correspondingly, the joint density can be expressed as $f(\bx,\by)\propto f_1(\bx)f_2(\by)\exp\{\phi(\bx,\by)\}$ for a function $f_1(\bx)$, so that the dependence between $\bX$ and $\bY$ is captured entirely by the interaction function $\phi$. The NETM thus provides a fully nonparametric, symmetric, and distribution-free representation of conditional dependence, in which the structure of the conditional density is learned from the data rather than imposed by a parametric baseline or an index assumption.

We estimate the interaction function $\phi$ by maximizing a profile empirical likelihood over a nonparametric sieve, with DNN sieves as the leading case. This framework offers a unified treatment of a wide range of inferential targets, including the mutual density ratio and mutual information, conditional features such as means, quantiles, and distribution functions, as well as tests of independence between $\bX$ and $\bY$. The main contributions of this paper are summarized as follows.

\bit
\item
{\it A unified conditional density representation with identifiability.}
We propose a unified nonparametric representation of the conditional density of $\bY$ given $\bX$ in which the dependence between the two random vectors is captured entirely by an interaction function $\phi(\bx,\by)$, and we establish identifiability of this representation under a suitable normalization condition. This representation provides a common and principled framework from which a broad class of quantities describing the relationship between $\bX$ and $\bY$,
including conditional distributions, mutual density ratios, and conditional features, can be derived, thereby unifying problems that have traditionally been treated separately.

\item
{\it Nonparametric likelihood inference and new empirical-process tools.}
Building on this representation, we develop a nonparametric likelihood method for estimating the unknown components and establish non-asymptotic error bounds for the resulting maximum likelihood estimators. The analysis is nontrivial: the profile empirical likelihood weights are defined implicitly through the observed responses, so the resulting profile likelihood is a nonstandard, sample-dependent criterion that admits neither a closed-form expression nor a readily tractable population counterpart. To address this difficulty, we construct a second-order approximation to the profile likelihood via an i.i.d. empirical average, which is amenable to conventional empirical process theory. The construction is itself highly challenging and hinges on an exponential tail bound for the supremum of a canonical U-process under a uniform entropy condition. This bound is, to the best of our knowledge, new to the U-process literature and constitutes the key technical contribution of this work.

\item
{\it A unified framework for multiple inferential targets.}
We show that the proposed likelihood framework yields a unified approach to three fundamental problems concerning the relationship between random vectors: estimation of conditional features, estimation of the mutual density ratio and mutual information, and testing for independence.
For conditional features, we develop a general theory for conditional $M$-functionals that covers conditional means, conditional quantiles, and conditional distribution functions as special cases; the resulting estimators achieve the same non-asymptotic rate as the underlying interaction estimator, while requiring regularity conditions only on \eqref{NETM} rather than separate smoothness assumptions for each conditional feature. For the mutual density ratio, we establish a non-asymptotic error bound showing that the estimator inherits the convergence rate of the estimated interaction function; as a direct consequence, the mutual information estimator attains the same order of convergence rate. For independence testing, we establish that independence is equivalent to the absence of the interaction component in the representation, and we construct a finite-sample valid permutation test based on the profile likelihood ratio statistic, which targets precisely the parameter whose vanishing characterizes independence.

\item
{\it Deep neural network sieves and adaptation to intrinsic structure.}
We further investigate the proposed estimators under DNN sieves. By exploiting the approximation properties of hierarchical composition function classes, we derive explicit non-asymptotic rates for the conditional feature, mutual density ratio,  and mutual information estimators. The resulting convergence rates depend on the intrinsic structural complexity of the interaction function rather than explicitly on the ambient dimension, demonstrating that the proposed framework can mitigate the curse of dimensionality under appropriate low-dimensional structural assumptions and is therefore well suited to high-dimensional, complex data.
\eit

\subsection{Related work}
\label{sec-related}
	{\it Conditional density and distribution estimation.}
	Nonparametric estimation of conditional densities and distributions has a long history.
	For conditional density estimation, kernel-based estimators were studied by
	\cite{rosenblatt1969} and \cite{hall2004}, and bandwidth selection by
	cross-validation is treated in \cite{hall2004}; these methods, however,
	suffer from the curse of dimensionality when $\bX$ is even moderately high-dimensional.
	Structured alternatives include conditional density estimators based on basis expansions
	\citep{izbicki2016, izbicki2017}, the boosting approach LinCDE of \cite{GaoHastie2022},
	which models the conditional density through a flexible exponential-family representation
	combined with gradient boosting, as well as neural conditional density models such as
	mixture density networks \citep{bishop1994} and normalizing flows \citep{papamakarios2021}. More recently, several methods have been developed to learn conditional distributions directly.
	\cite{zhou2023deep}, \cite{shen2025engression}, and \cite{song2026wasserstein} proposed deep generative approaches based on conditional generators and different distribution-matching criteria, while \cite{cevid2022distributional} developed distributional random forests using MMD-based splitting and forest-induced weights. Our exponential tilt representation \eqref{NETM} is close in spirit to LinCDE in that the
	log conditional density is modeled additively in a baseline term and an interaction term,
	but we leave the baseline fully nonparametric and profile it out by empirical likelihood
	rather than parameterizing it, which avoids the need to specify the baseline family.
	
	{\it Mutual density ratio and mutual information estimation.}
	The mutual density ratio (MDR), defined as the joint-to-product density ratio, provides a pointwise characterization of dependence between random variables. Although general density-ratio estimation has been extensively studied
	\citep[see, e.g.,][]{sugiyama2012}, more recent work has considered direct
	estimation of the MDR and closely related ratios.
	\cite{matsushita2023estimating} studied direct least-squares estimation of the product-of-marginals-to-joint density ratio, the reciprocal of the MDR, using series approximation.
	\cite{Han2025} developed a Bregman-divergence framework for MDR estimation, using deep neural networks to approximate the log ratio and establishing non-asymptotic error bounds. Under the least-squares loss, this approach is closely related to the density-ratio fitting method for pointwise dependency of \cite{Tsai2020}.
	Mutual information (MI) \citep{shannon1948, cover2006}, given by the expectation of the log MDR, is a canonical scalar measure of dependence. Representative estimators include nearest-neighbor-based methods \citep{kraskov2004, gao2018demystifying, berrett2019entropy} and neural variational approaches such as MINE \citep{belghazi2018}.
	Our approach estimates the MDR through a likelihood-based estimate of the interaction function rather than by minimizing a divergence, while simultaneously yielding a properly normalized conditional density. The resulting plug-in estimator of MI inherits the convergence rate of the estimated interaction function.
	
{\it Independence testing.}
	Testing $\bX\perp\bY$ is a classical problem. Modern nonparametric tests include distance correlation \citep{szekely2007, szekely2013}, the Hilbert--Schmidt independence criterion \citep{gretton2005, gretton2008}, rank-based and copula-based tests \citep{deb2023, shi2022}, and tests based on mutual information estimates \citep{berrett2019}. Permutation calibration is standard for such statistics, and it is well known that a permutation test based on any statistic is finite-sample valid under exchangeability \citep{lehmann2005}. \cite{Han2025} also proposed a permutation test based on their neural mutual information estimator. In contrast, our test uses the profile likelihood ratio statistic, which compares the fitted interaction function against $\phi\equiv0$ and therefore targets precisely the parameter whose vanishing characterizes independence, yielding a test that is both interpretable and tightly connected to the underlying dependence structure.
	
{\it Deep neural network theory.}
	The statistical theory of DNN estimators has advanced rapidly. Approximation results for ReLU networks \citep{yarotsky2017, shen2022} combined with VC-dimension bounds \citep{bartlett2019nearly} yield convergence rates for nonparametric regression \citep{ schmidtHieber2020}, classification \citep{kim2021}, quantile regression \citep{ shen2024}, survival analysis \citep{Zhong2022}, and mutual density-ratio estimation \citep{Han2025}. A recurring theme is that when the target function has a hierarchical composition structure, the rate depends on the intrinsic dimension of the composition rather than the ambient dimension \citep{bauer2019, schmidtHieber2020}. We adopt the hierarchical composition function (HCF) classes of \cite{yan2025deep}, which allow a possibly different smoothness at each node of the composition tree, and derive rates for all  inferential targets under a single HCF assumption on $\phi_0$, thereby unifying the rate analysis across estimation, inference, and testing.
	
{\it U-processes.}
	A key technical ingredient of our analysis is an exponential tail bound for the supremum of a canonical U-process indexed by a function class. Uniform limit theory for U-processes was developed by \cite{nolan1987}, \cite{arcones1993} and \cite{sherman1994}; see \cite{delapena1999} for a comprehensive treatment. Concentration inequalities for canonical U-statistics of fixed kernels are available from \cite{gine2000} and \cite{houdre2003}, and for U-processes indexed by finite classes from \cite{clemencon2008}. Our bound is stated under a uniform entropy condition on the indexing class, which is exactly the form needed for sieve estimators, and we give a largely self-contained proof.

	\subsection{Organization}
	
	The rest of the paper is organized as follows. Section~\ref{sec-method} investigates the flexibility of the NETM, establishes its identifiability, and develops the profile empirical likelihood estimator. Section 3 establishes an exponential tail bound for the supremum of a canonical U-process under a uniform entropy condition, develops non-asymptotic theory for the proposed estimators under general sieve conditions, and proves finite-sample validity of the proposed independence test.  Section~\ref{sec-dnn} specializes the results to deep neural network sieves and hierarchical composition function classes. Section~\ref{section-simu} reports simulation studies, Section~\ref{sec-real-data} presents an analysis of the eye screening data, and Section~\ref{sec-discussion} concludes the paper. Proofs  and additional numerical details are provided in the Supplementary Material.

	\section{Methodology}
	\label{sec-method}
	
	Suppose that $\{(\bX_1, \bY_1), \ldots, (\bX_n, \bY_n)\}$ are $n$ independent
	and identically distributed (i.i.d.) observations from $(\bX, \bY)$.
	Let $\mathcal{X}$ and $\mathcal{Y}$
	denote the supports of $\bX$ and $\bY$, respectively.
	We express the conditional density of $\bY$ given $\bX$ as
	the NETM in \eqref{NETM}. Our goals are to
	estimate the interaction function $\phi(\bx,\by)$ and
	the conditional distribution function
	\[
	F_{Y\mid X}(\by \mid \bx) = \int_{ \mathcal{Y}} f(\bv\mid\bx) I(   \bv\leq \by)\, d\bv
	=
	\frac{ \int_{ \mathcal{Y} } f_2(\bv)\exp\{\phi(\bx,\bv)\}I(\bv\leq \by)\, d\bv   }{\int_{ \mathcal{Y} } f_2(\bv)\exp\{\phi(\bx,\bv)\}\,d\bv},
	\]
	and to make inference about certain features of $(\bX, \bY)$.
	
	\subsection{Flexibility of the NETM}
	
	Before presenting our estimation procedure, we examine the flexibility of the
	NETM \eqref{NETM}. Under this model, the joint density of $(\bX,\bY)$ admits
	the representation
	\ba
	\label{model-joint-density}
	f(\bx,\by)
	=
	\frac{
		f_1(\bx) f_2(\by) \exp\{\phi(\bx,\by)\}
	}{
		\iint_{\mathcal X\times\mathcal Y}
		f_1(\bu)f_2(\bv)\exp\{\phi(\bu,\bv)\}\,d\bu\,d\bv
	},
	\ea
	for any $(\bx, \by) \in \mathcal{X}\times \mathcal{Y}$, where $f_1(\bx)$ is a baseline density on $\mathcal X$. Indeed, one may take $f_{1}(\bx)\propto f_{x}(\bx)/(\int_{ \mathcal{Y} } f_2(\bv)\exp\{\phi(\bx,\bv)\}\,d\bv)$, where $f_{x}$ is the marginal density of $\bX$. In other words, given a nonparametric model
	\eqref{NETM} for $f(\by\mid\bx)$, the joint density $f(\bx, \by)$ admits a
	fully nonparametric representation \eqref{model-joint-density}.
	
	Model \eqref{model-joint-density} can be viewed as a copula-like decomposition.
	Recall that, by Sklar's theorem \citep{sklar1959, nelsen2006}, classical copula
	models separate marginal behavior from dependence through a copula after
	probability-integral transformations. In \eqref{model-joint-density}, an analogous
	separation is achieved directly on the original scale and additively on the log scale:
	\[
	\log f(\bx,\by)
	=
	\log f_1(\bx)+\log f_2(\by)+\phi(\bx,\by)-\log Z_\phi,
	\]
	where
	\(
	Z_{\phi}
	=
	\iint_{\mathcal{X}\times\mathcal{Y}}
	f_1(\bu)f_2(\bv)\exp\{\phi(\bu,\bv)\}\,d\bu\,d\bv.
	\)
	Thus, $\phi(\bx,\by)$ captures the dependence between $\bX$ and $\bY$, while
	$f_1$ and $f_2$ serve as baseline components for the two random vectors.
	Indeed, $\bX$ and $\bY$ are independent if and only if $\phi$ is additively
	separable. Under the normalization introduced in
	Section~\ref{sec-identifiability}, this is equivalent to $\phi\equiv0$
	(see Section~\ref{sec-application}). Unlike a classical copula construction,
	the representation operates directly on $(\bx,\by)$ without marginal
	transformations and applies without modification when $\bX$ and $\bY$ are
	multivariate.
	
	Denote the marginal density of $\bY$ as $f_{y}$. Integrating \eqref{model-joint-density} with respect to $\by$ and $\bx$,
	respectively, yields
	\ba
	\label{marginal-densities}
	f_x(\bx) = \frac{f_1(\bx) \int f_2(\bv) \exp\{ \phi(\bx, \bv) \}\, d\bv}{Z_\phi},
	\;
	f_y(\by) = \frac{f_2(\by) \int f_1(\bu) \exp\{ \phi(\bu, \by) \}\, d\bu}{Z_\phi}.
	\ea
	Thus, the marginal density of $\bX$ is obtained by reweighting the baseline
	density $f_1(\bx)$ by the factor $\int f_2(\bv)\exp\{\phi(\bx,\bv)\}\,d\bv$,
	and analogously for $\bY$. When $\phi\equiv 0$, these reweighting factors are
	constant, so that $f_x$ is proportional to $f_1$ and $f_y$ is proportional to
	$f_2$. For this reason, we refer to $f_1$ and $f_2$ as \emph{baseline
		densities} rather than marginal densities. Notably, dividing \eqref{model-joint-density} by the marginal density
	$f_x(\bx)$ in \eqref{marginal-densities} immediately recovers the NETM
	\eqref{NETM}. Thus, given the marginal distribution of $\bX$, the NETM \eqref{NETM} and the joint representation
	\eqref{model-joint-density} are equivalent.
	
	\subsection{Identifiability}
	\label{sec-identifiability}
	
	Without additional restrictions on $f_1(\bx)$, $f_2(\by)$ and $\phi(\bx, \by)$, models \eqref{model-joint-density}
	and \eqref{NETM} are clearly not identifiable: for instance, one can multiply $f_1$ by any positive function $a(\bx)$ and subtract $\log a(\bx)$ from $\phi$ without changing $f(\bx,\by)$. Let $\Psi_{1}$ and $\Psi_{2}$ be fixed probability measures on $\mathcal{X}$ and $\mathcal{Y}$, respectively. Define
	\bas
	\mathscr{P}:=\left\{\phi:\mathcal{X}\times\mathcal{Y}\mapsto\mathbb{R}\bigg| 0 =\int_{\mathcal X} \phi(\bu,\by)\,d\Psi_1(\bu)= \int_{\mathcal Y}\phi(\bx,\bv)\,d\Psi_2(\bv) \right\}.
	\eas
	
	\begin{prop}
		\label{prop-identifiability}
		Suppose $\int f_1(\bx)d\bx = \int f_2(\by)d\by =1$ and $\phi\in\mathscr{P}$ in model \eqref{model-joint-density}. Then $(f_{1}(\cdot), f_2(\cdot), \phi(\cdot))$ are identifiable from model \eqref{model-joint-density}.
	\end{prop}
	
	For every bounded measurable function
	$f:\mathcal X\times\mathcal Y\to\mathbb R$, define the identifying
	projection
	\[
	\begin{aligned}
		(\mathcal C f)(\bx,\by)
		&=
		f(\bx,\by)
		-
		\int_{\mathcal X}f(\bu,\by)\,d\Psi_1(\bu)
		-
		\int_{\mathcal Y}f(\bx,\bv)\,d\Psi_2(\bv)
		+
		\iint_{\mathcal X\times\mathcal Y}
		f(\bu,\bv)\,d\Psi_1(\bu)d\Psi_2(\bv).
	\end{aligned}
	\]
	The following construction shows that a normalized triple always exists. Suppose that $f_{1}$, $f_2$ and $\phi$ satisfy model \eqref{model-joint-density}.
	Let
	$
	\phi'(\bx, \by) = (\mathcal C \phi )(\bx,\by).
	$
	Set
	\bas
	f_1'(\bx) =  \frac{f_1 (\bx) \exp\left\{ \int  \phi(\bx, \by) d\Psi_{2}(\by) \right\}}{\int f_1 (\bx) \exp\left\{ \int  \phi(\bx, \by) d\Psi_{2}(\by) \right\}d\bx}, \;
	f_2'(\by) =  \frac{f_2 (\by) \exp\left\{ \int  \phi(\bx, \by) d\Psi_{1}(\bx) \right\}}{\int f_2 (\by) \exp\left\{ \int \phi(\bx, \by) d\Psi_{1}(\bx)  \right\} d\by}.
	\eas
	It can be verified that $\int f_{1}'(\bx)d\bx = \int f_2'(\by)d\by = 1$ and $\phi'\in\mathscr{P}$. Since
	\bas
	\frac{f_1(\bx) f_2(\by) \exp\{ \phi(\bx, \by) \}  }{
		\iint  f_1(\bu) f_2(\bv) \exp\{ \phi(\bu, \bv) \} d\bu d\bv}
	=
	\frac{f_1'(\bx) f_2'(\by) \exp\{ \phi'(\bx, \by) \}  }{
		\iint  f_1'(\bu) f_2'(\bv) \exp\{ \phi'(\bu, \bv) \} d\bu d\bv},
	\eas
	the triple $(f_{1}'(\bx), f_2'(\by), \phi'(\bx, \by))$ also satisfies \eqref{model-joint-density} and, by Proposition~\ref{prop-identifiability}, is the unique normalized representative. In practice, $\Psi_1$ and $\Psi_2$ may be taken to be the empirical distributions of $\bX$ and $\bY$.
	
	\subsection{Estimation of $\phi$}
	
	By \eqref{NETM},
	the conditional likelihood function is
	\bas
	\prod_{i=1}^n \frac{  f_2(\bY_i) \exp\{ \phi(\bX_i, \bY_i) \}  }{
		\int_{\mathcal{Y}}   f_2(\bv) \exp\{ \phi(\bX_i, \bv) \}   d\bv}.
	\eas
	Following the principle of empirical likelihood \citep{owen1988, owen2001, qin1994}, we
	model the baseline distribution function $F_2(\by) =  \int_{\mathcal{Y}}  f_2(\bv)  I(\bv\leq \by) d\bv$
	by a step function $F_2(\by) = \sum_{i=1}^n p_i I(\bY_i\leq \by) $, where $p_i\geq 0$
	and $\sum_{i=1}^n  p_i =1$. Then the conditional log-likelihood function
of ${\bm p} = (p_{1},\cdots, p_{n})$  and $\phi$ given $\{\bX_1, \ldots, \bX_n\}$ is
	\ba
	\label{definition-conditional-likelihood}
	\ell_{c}({\bm p}, \phi)  = \sum_{i=1}^n \bigg\{  \log(p_i) + \phi(\bX_i, \bY_i)  - \log \bigg[ \sum_{j=1}^n  p_j \exp\{\phi(\bX_i, \bY_j)\}\bigg] \bigg\}.
	\ea

	The marginal likelihood function based only on $\{\bX_1, \ldots, \bX_n\}$ is
	\bas
	\prod_{i=1}^n \frac{f_1(\bX_i)  \int_{\mathcal{Y}}  f_2(\by) \exp\{ \phi(\bX_i, \by) \}  d\by }{
		\iint_{\mathcal{X}\times\mathcal{Y}}   f_1(\bu) f_2(\bv) \exp\{ \phi(\bu, \bv) \} d\bu d\bv}.
	\eas
	Similarly, we
	model the baseline distribution function $F_1(\bx) =  \int_{\mathcal{X}} f_1(\bu) I(\bu\leq \bx) d\bu$
	by a step function $F_1(\bx) = \sum_{i=1}^n q_i I(\bX_i\leq \bx) $, where $q_i\geq 0$
	and $\sum_{j=1}^n  q_j =1$.
	Then the log-likelihood becomes
	\bas
	\ell_m({\bm q}, \phi, \bp) = \sum\limits_{i=1}^{n}\log\{q_{i}w(\bX_{i};\phi, \bp)\} - n\log\bigg\{\sum\limits_{j=1}^{n}q_{j}w(\bX_{j};\phi, \bp)\bigg\},
	\eas
	where ${\bm q} = (q_{1}, \cdots, q_{n})$ and $w(\bx;\phi, \bp) = \sum_{i=1}^{n}p_{i}\exp\{\phi(\bx, \bY_{i})\}$. Denote by $\mathscr{F}_n^{(0)}$ a nonparametric sieve space approximating $\phi$, and let $\mathscr{F}_{n} = \{(\mathcal{C}\phi):\phi\in\mathscr{F}_n^{(0)}\}$.
	We estimate $(\bp, \bq, \phi)$ by the maximum likelihood estimator (MLE)
	\ba
	\label{defi-MLE-full-likelihood}
	(\widehat{\bp}, \widehat{\bq}, \widehat{\phi}) = \operatorname*{argmax}_{\bp, \bq\in\mathcal{P}, \phi\in\mathscr{F}_n} \{ \ell_{c}(\bp, \phi)+\ell_{m}(\bq, \phi, \bp)\},
	\ea
	where $\mathcal{P}: = \{\bp:\sum_{j=1}^{n}p_{j}=1, p_{j}\geq 0\}$.
	Once $\widehat{\bp}$ and $\widehat{\bq}$ are available,
	our MLEs for $F_1$ and $F_2$ are
	\(
	\widehat{F}_{1}(\bx) = \sum_{i=1}^{n}\widehat{q}_{i}I(\bX_{i}\leq \bx)\) and \(\widehat{F}_{2}(\by) = \sum_{i=1}^{n}\widehat{p}_{i}I(\bY_{i}\leq \by),
	\) respectively.
	
	We propose to solve \eqref{defi-MLE-full-likelihood} by profiling.
	For given $\bp$ and $\phi\in\mathscr{F}_n$, the optimal $\bq$ that maximizes \eqref{defi-MLE-full-likelihood}
	is $\widehat{\bq}(\bp, \phi)$, whose $i$th component is
	$
	\widehat{q}_{i}(\bp, \phi) = \frac{1/w(\bX_i; \phi, \bp)}{ \sum_{j=1}^n  1/w(\bX_j; \phi, \bp) }.
	$
	This leads to the profile log-likelihood function of $(\bp, \phi)$
	\bas
	\ell_{\text{profile},1}(\bp, \phi) = \ell_{c}(\bp, \phi)+\ell_{m}(\widehat{\bq}(\bp, \phi), \phi, \bp) =  \ell_{c}(\bp, \phi) - n\log n.
	\eas
	Given $\phi$,
	setting the partial derivative of $\ell_{\text{profile},1}(\bp, \phi)$ with respect to $p_i$ to zero gives
	\ba  \label{original-empirical-score-equation}
	p_r = \left\{ \sum_{i=1}^n   \frac{     \exp\{\phi(\bX_i, \bY_r)\} }{ \sum_{j=1}^n  p_j \exp\{\phi(\bX_i, \bY_j) \}}
	\right\}^{-1},~~ r = 1,\cdots,n.
	\ea
	This is a fixed-point equation of the same type as that arising in the proportional likelihood ratio model \citep{luo2012, davidov2013}, and may be solved by fixed-point iteration initialized at $p_i=1/n$.
	Denote the resulting estimator by $\widehat{\bp}(\phi): = (\widehat p_1 (\phi), \cdots, \widehat p_n (\phi))$, which satisfies $\sum_{i=1}^{n}\widehat p_{i}(\phi) = 1$.
	Substituting $\widehat{\bp}(\phi)$ back into $\ell_{\text{profile},1}(\bp, \phi) $, we obtain
	the profile log-likelihood of $\phi$
	\bas
	\ell_{\text{profile},2}(\phi)   =
	\sum_{i=1}^n \bigg\{   \log\{ \widehat p_i(\phi)\} + \phi(\bX_i, \bY_i)  - \log \bigg[ \sum_{j=1}^n  \widehat p_j (\phi) \exp\{\phi(\bX_i, \bY_j)\}\bigg] \bigg\},
	\eas
	where a constant not depending on $\phi$ is ignored.
	It then follows that
	\ba
	\label{defi-MLE-profile-likelihood-2}
	\widehat{\phi} = \operatorname*{argmax}_{\phi\in\mathscr{F}_n}\ell_{\text{profile},2}(\phi).
	\ea
	Once $\widehat \phi$ is obtained,
	we have $\widehat{\bp} \equiv   (\widehat p_1, \ldots, \widehat p_n) = \widehat{\bp}(\widehat \phi)$ and
	$\widehat{\bq} \equiv (\widehat q_1, \ldots, \widehat q_n)  = \widehat{\bq}(\widehat \bp, \widehat \phi)$.
	
	\subsection{Estimation of $F_{Y\mid X}(\by \mid \bx)$ and conditional features}
	\label{sec-conditional-distribution}
	
	With the MLEs $\widehat \phi$, $\widehat F_1$ and $\widehat F_2$,
	our MLE for $F_{Y\mid X} (\by\mid\bx)$ is
	\ba
	\label{estimated-conditional-distribution}
	\widehat F_{Y\mid X} (\by\mid\bx)
	=
	\frac{
		\displaystyle
		\int_{\mathcal{Y}} I(\bv\leq\by)
		\exp\{\widehat\phi(\bx,\bv)\}
		\,d\widehat F_2(\bv)
	}{
		\displaystyle
		\int_{\mathcal{Y}}
		\exp\{\widehat\phi(\bx,\bv)\}
		\,d\widehat F_2(\bv)
	}
	=
	\sum_{i=1}^n
	\widehat\omega_i(\bx)
	I(\bY_i\leq\by),
	\ea
	where
	\ba
	\label{estimated-conditional-weights}
	\widehat\omega_i(\bx)
	=
	\frac{
		\widehat p_i
		\exp\{\widehat\phi(\bx,\bY_i)\}
	}{
		\displaystyle
		\sum_{j=1}^n
		\widehat p_j
		\exp\{\widehat\phi(\bx,\bY_j)\}
	},
	\qquad
	i=1,\ldots,n.
	\ea
	It can be seen that $\widehat F_{Y\mid X} (\cdot\mid\bx)$ is a weighted empirical distribution of the observed responses, with weights depending on the covariate value $\bx$.
	
	Given an estimator of $F_{Y\mid X}(\by \mid \bx)$, we can construct
	estimators for a variety of conditional features. To fix ideas, let
	\(
	T:\mathcal{Y}\to\mathbb{R}
	\)
	be a given measurable scalar-valued function, let
	$\Theta\subseteq\mathbb{R}$ be a parameter space, and let
	\(
	\ell:\mathbb{R}\times\Theta\to\mathbb{R}
	\)
	be a measurable loss function. We consider the conditional feature
	\bas
	\vartheta_0(\bx)
	\in
	\operatorname*{argmin}_{\vartheta\in\Theta}
	\e\left\{
	\ell(T(\bY),\vartheta)
	\mid\bX=\bx
	\right\}.
	\eas
	When $\bY$ is univariate, a natural choice is
	$T(\by)=\by$; when $\bY$ is multivariate, $T$ may select a particular
	coordinate or any other scalar summary of interest.
	This formulation is very general and encompasses many commonly used
	parameters as special cases. Prominent examples include the conditional
	mean, obtained with the squared loss
	$\ell(U,\vartheta)=(U-\vartheta)^2$; the conditional $\tau$-quantile,
	obtained with the check loss
	$\ell(U,\vartheta)=\rho_\tau(U-\vartheta)$, where
	\(\rho_\tau(u)=u\{\tau-I(u<0)\}\); and the conditional distribution
	function, obtained with the squared loss
	$\ell(U,\vartheta)=(U-\vartheta)^2$ and
	$U=I(\bY\leq\by)$.
	
	Given the estimate in \eqref{estimated-conditional-distribution},
	a natural plug-in estimator of $\vartheta_0(\bx)$ is the weighted conditional
	$M$-estimator
	\ba
	\label{proposed-condiitonal-mean}
	\widehat\vartheta(\bx)
	\in
	\operatorname*{argmin}_{\vartheta\in\Theta}
	\sum_{i=1}^n
	\widehat\omega_i(\bx)
	\ell( U_i,\vartheta),
	\ea
	where $U_i=T(\bY_i)$ and $\widehat\omega_i(\bx)$ is defined in
	\eqref{estimated-conditional-weights}.
	
	\section{Non-asymptotic guarantees}
	\label{sec-theory}
	
	In this section, we establish non-asymptotic error bounds for the proposed estimators
	$\widehat\phi$, $\widehat F_{Y\mid X}(\by\mid\bx)$, and $\widehat\vartheta(\bx)$.
	Let $P_0$ denote the true joint distribution of $(\bX,\bY)$, and let $P_X$
	and $P_Y$ be the corresponding marginal distributions. For any measurable
	functions $a(\bx)$, $b(\by)$, and $c(\bx,\by)$, define
	$
	P_X(a)=\int a(\bx)\, dP_X(\bx)$, $
	P_Y(b)=\int b(\by)\, dP_Y(\by)$, $
	P_0(c)=\int c(\bx,\by)\, dP_0(\bx,\by),
	$
	and
	$
	\|a\|_{L_2(P_X)} = \{P_X(a^2)\}^{1/2}$,
	$\|b\|_{L_2(P_Y)} = \{P_Y(b^2)\}^{1/2},
	$
	$
	\|c\|_{L_2(P_XP_Y)} =
	\{\int_{\mathcal{X}\times\mathcal{Y}}
	c(\bx,\by)^2\, dP_X(\bx)\, dP_Y(\by) \}^{1/2}.
	$
	For any $h:\mathcal{X}\times\mathcal{Y}\to\mathbb{R}$, let $\|h\|_\infty$
	denote its supremum norm. We use $P$, $P_x$, and $P_y$ to denote generic
	probability measures on $\mathcal{X}\times\mathcal{Y}$, $\mathcal{X}$, and
	$\mathcal{Y}$, respectively. For any function class $\mathcal{F}$ defined on
	the metric space $(\mathcal{X}\times\mathcal{Y}, \|\cdot\|)$, the
	$\epsilon$-cover number $N(\epsilon, \mathcal{F}, \|\cdot\|)$ is defined as
	the smallest number of balls of radius $\epsilon$ whose union contains
	$\mathcal{F}$.
	Let $f_0(\bx,\by)$ denote the true joint density of $(\bX,\bY)$; the
	quantities $f_{x,0}$, $f_{y,0}$, $f_{1,0}$, $f_{2,0}$, $\phi_0$, and
	$F_{Y\mid X,0}$ are defined analogously. Denote by $\Pr(A)$ the probability of a random event $A$.

	\subsection{Error bound for $\widehat \phi$}

Establishing a non-asymptotic error bound for $\widehat\phi$ is nontrivial, as the profile likelihood in \eqref{defi-MLE-profile-likelihood-2} admits neither a closed-form expression nor a readily tractable population counterpart. To address this difficulty, we construct a second-order approximation to the profile likelihood by an i.i.d. empirical average, which is amenable to conventional empirical process theory. The construction is itself highly challenging and hinges on an exponential tail bound for the supremum of a canonical U-process under a uniform entropy condition (Theorem \ref{thm-canonical-U-process}); this bound is new to the U-process literature. The theorem is moreover of independent interest, and our proof of it is largely self-contained, invoking only a small number of standard auxiliary inequalities.

For a measurable space $\mathcal{Z}$ with distribution $P_{z}$, a symmetric kernel \(u:\mathcal Z^2\to\mathbb R\) is called
\(P_z\)-canonical if
$
P_z\{u(\bz,\bZ)\}
=
\int u(\bz,\bz')\,dP_z(\bz')
=
0
$
for \(P_z\)-almost every \(\bz\in\mathcal Z\).
We take $\bZ = (\bX^\T, \bY^\T)^\T$  when applying this theorem in our setting. 

\begin{theorem}
	\label{thm-canonical-U-process}
	Let $\bZ_1,\ldots,\bZ_n$ be i.i.d. random vectors taking values in
	a measurable space $\mathcal Z$ with  distribution $P_z$.
	Let $\mathcal U_n$ be a pointwise measurable class of symmetric,
	$P_z$-canonical kernels \(
	u:\mathcal Z^2\longmapsto\mathbb R.
	\)
	Suppose that $\mathcal U_n$ has a
	constant envelope $B<\infty$ and that
	\[
	\sup_Q
	\log N\left(
	\epsilon B,\mathcal U_n,L_2(Q)
	\right)
	\le
	C_{\rm E}\tilde V_n\log(\tilde A_n/\epsilon),
	\qquad
	0< \epsilon\le 1
	\]
for some  deterministic sequences $\tilde V_n\ge1$ and $\tilde A_n\ge e$, and a positive constant $C_{\rm E}$,
	where the supremum is taken over all finitely discrete probability
	measures $Q$ on $\mathcal Z^2$. Then there exists a constant $\tilde C>0$ depending only on $C_{\rm E}$ such that
	\bas
	\Pr\bigg[
	\sup_{u\in\mathcal U_n}
	\bigg|
	\frac1{n(n-1)}
	\sum_{i\ne j}u(\bZ_i,\bZ_j)
	\bigg|
	>
	\tilde CB
	\frac{
		\tilde V_n\log(\tilde A_n)+t
	}{
		n
	}
	\bigg]
	\le
	e^{-t}, \quad \forall n\geq 2, \; t\geq 1.
	\eas
\end{theorem}

	Recall that
	$\mathscr{F}_{n} = \{(\mathcal{C}\phi):\phi\in\mathscr{F}_n^{(0)}\}$,
	where $\mathscr{F}_n^{(0)}$ is a nonparametric sieve space approximating $\phi$.
	Define the $\mathcal X$-slice class
	\(
	\mathscr F_{n,X}^{\mathrm{sl}}
	=
	\left\{
	\bx\mapsto \phi(\bx,\by):
	\phi\in\mathscr F_n,\ \by\in\mathcal Y
	\right\}
	\)
	and the $\mathcal Y$-slice class
	\(
	\mathscr F_{n,Y}^{\mathrm{sl}}
	=
	\left\{
	\by\mapsto \phi(\bx,\by):
	\phi\in\mathscr F_n,\ \bx\in\mathcal X
	\right\}.
	\)
	Our error bound for $\widehat \phi$ relies on the following assumptions.
	
	\begin{assumption}
		\label{ass-bounded-sieves}
		There exists a finite constant $D>0$ such that
		\(
		\sup_{\phi\in\mathscr F_n}\|\phi\|_\infty\le D
		\).
	\end{assumption}
	
	\begin{assumption}
		\label{ass-entropy-full-sieves}
		There exist deterministic sequences $V_n\ge 1$ and $A_n\ge e$ such that, for all
		$0<\epsilon\le 1$, (i)
		\(
		\sup_{P} \log \{N (\epsilon D,\mathscr F_n,L_2(P) )\}
		\le
		V_n\log(A_n/\epsilon),
		\) (ii)
		\(
		\sup_{P_{x}}
		\log \{N (\epsilon D,\mathscr F_{n,X}^{\rm sl},L_2(P_{x}) )\}
		\le
		V_n\log(A_n/\epsilon),
		\)
		(iii)
		\(
		\sup_{P_{y}}
		\log \{N (\epsilon D,\mathscr F_{n,Y}^{\rm sl},L_2(P_y) )\}
		\le
		V_n\log(A_n/\epsilon),
		\)
		where the suprema are taken over all probability measures on the corresponding
		spaces.
	\end{assumption}
	
	\begin{assumption}
		\label{ass-true-function-identification}
		(i)	 $\phi_0\in\mathscr{P}$ and
		\(
		\|\phi_0\|_\infty\le D
		\).
		(ii)
		\(
		\Psi_1\ll P_X,~
		\Psi_2\ll P_Y,
		\)
		and there exist constants $0<K_1,K_2<\infty$ such that
		\(
		\left\|
		d\Psi_1/dP_X
		\right\|_{L_2(P_X)}
		\le K_1\) and
		\(
		\left\|
		d\Psi_2/dP_Y
		\right\|_{L_2(P_Y)}
		\le K_2.
		\)
	\end{assumption}
	
	\begin{assumption}
		\label{ass-approximation}
		There exists a deterministic sequence
		$ a_n\ge0$ and an element
		$\phi_n^\star\in\mathscr F_n$ such that
		\(
		\|\phi_n^\star - \phi_0\|_{L_{2}(P_{X}P_{Y})}
		\le
		2 a_n.
		\)
	\end{assumption}
	
	Assumption \ref{ass-bounded-sieves} is trivially satisfied. Assumption \ref{ass-entropy-full-sieves} imposes complexity conditions on the nonparametric sieves; it holds, in particular, for sieves with controlled VC dimension or pseudo-dimension (see Definitions 1--2 in \citealp{bartlett2019nearly}). Assumption \ref{ass-true-function-identification} is standard. Assumption~\ref{ass-approximation} provides a deterministic upper bound on the sieve approximation error.
	Define $
	a_n^\circ \equiv \inf_{\phi\in\mathscr F_n}\|\phi-\phi_0\|_{L_2(P_XP_Y)} \geq 0
	$. If
	$
	a_n^\circ   > 0
	$
	or the infimum is attained, one may take \(a_n = a_n^\circ\).

	For \(\phi_1,\phi_2:\mathcal{X}\times\mathcal{Y}\to\mathbb{R}\), define
	$
	d(\phi_1,\phi_2)=\|\phi_1-\phi_2\|_{L_2(P_XP_Y)}.
	$
	Denote
	$
	\delta_n=\left\{V_n\log(nA_n)/n\right\}^{1/2},
	$
	where \(A_n\) and \(V_{n}\) are as in Assumption \ref{ass-entropy-full-sieves}.
	The quantity $\delta_n$ will be used repeatedly hereafter.
	
	\begin{theorem}
		\label{prop-rate-hat-phi}
		Suppose that Assumptions \ref{ass-bounded-sieves}-\ref{ass-approximation} hold.
		Then there exists a constant $C_\star>0$, depending only on $D,K_1$, and $K_2$, such that
		$$
		\Pr\left( d(\widehat\phi, \phi_0)
		\le
		C_\star
		\{
		a_n
		+
		\sqrt{\log(6/\varepsilon)}\,\delta_n
		\} \right)
		\geq 1-\varepsilon,$$
		for every $\varepsilon$ satisfying $0 < \varepsilon <1$ and
		\(
		\max\{\log\left(6/\varepsilon\right), 2\}\le n\).
	\end{theorem}
	
	Theorem~\ref{prop-rate-hat-phi} indicates that the error bound for the proposed estimator of the interaction function consists of two components: a stochastic error of order $\delta_{n}$ and a sieve approximation error of order $a_{n}$. Indeed, we can also show that for any bounded functions $a(\bx)$ and $b(\by)$, the bounds of
	the errors $ |\int b(\by)d\widehat{F}_{2}(\by) - \int f_{2,0}(\by)b(\by)d\by |$ and $ |\int a(\bx)d\widehat{F}_{1}(\bx) - \int f_{1,0}(\bx)a(\bx)d\bx |$ are of the same order as that of $d(\widehat{\phi},\phi_{0})$. Since $f_{1,0}$ and $f_{2,0}$ are only baseline measures, these results are not essential to the non-asymptotic results that follow and are therefore omitted.

	\subsection{Error bounds for the conditional feature estimator}
	\label{sec-conditional-M-estimation}
	
	We focus on loss functions that are convex in $\vartheta$,
	although our estimation procedure is applicable to general loss functions.
	For a loss function $\ell(u,\vartheta)$ convex in $\vartheta$, let
	\(
	\partial^{\rm s}_\vartheta\ell(u,\vartheta)
	\)
	denote the convex subdifferential with respect to $\vartheta$.
	For each $\vartheta\in\Theta$, let
	$\mathcal S_\ell(\vartheta)$ be the collection of all measurable
	functions $s(\cdot,\vartheta)$ satisfying
	\(
	s(u,\vartheta)
	\in
	\partial^{\rm s}_\vartheta\ell(u,\vartheta)
	\) for every \(u\).
	Define
	\(
	\mathcal S_\ell
	=
	\left\{
	u\mapsto s(u,\vartheta):
	\vartheta\in\Theta,\
	s(\cdot,\vartheta)\in\mathcal S_\ell(\vartheta)
	\right\}.
	\) For any fixed $\bx$, denote $M_{\bx}(\vartheta) = \e\left[
	\ell\{U,\vartheta\}
	\mid\bX=\bx
	\right]$, where $U =T(\bY)$.
	
	\begin{assumption}
		\label{ass-conditional-M-regularity}
		
		(i) $\Theta\subseteq\mathbb R$ is compact and convex,
		and for every $u\in\mathbb R$, the map
		\(
		\vartheta\longmapsto\ell(u,\vartheta)
		\)
		is lower semicontinuous and convex.
		(ii)
		There exist constants $B_\ell<\infty$, \(
		V_\ell\ge1\) and
		\(A_\ell\ge e,
		\) such that
		\(
		\sup_{s\in\mathcal S_\ell}\|s\|_\infty
		\le
		B_\ell,
		\) and
		\(
		\log N\left(
		\epsilon B_\ell,
		\mathcal S_\ell,
		L_2(Q)
		\right)
		\le
		V_\ell\log(A_\ell/\epsilon)
		\)
		for every probability
		measure $Q$ on $\mathbb R$ and $\epsilon\in(0,1]$.
		(iii)
		There exist constants
		\(
		r_\ell>0\) and \(\kappa_\ell>0\),
		such that
		\(
		\{\vartheta-\vartheta_0(\bx)\}v
		\ge
		\kappa_\ell
		|\vartheta-\vartheta_0(\bx)|^2 ,
		\) for any
		\(
		v\in\partial^{\rm s} M_{\bx}(\vartheta)
		\) and
		$P_X$-almost any $\bx$ and any
		\(
		\vartheta\in\Theta
		\)
		satisfying
		\(
		|\vartheta-\vartheta_0(\bx)|
		\le r_\ell
		\).
		(iv)
		\(
		\inf_{\bx}
		\inf_{\{
			\vartheta\in\Theta:
			|\vartheta-\vartheta_0(\bx)|=r_\ell
			\}}
		\left[
		M_{\bx}(\vartheta)
		-
		M_{\bx}\{\vartheta_0(\bx)\}
		\right]
		\ge
		\Delta_\ell
		\)
		for some
		$\Delta_\ell>0$.
		
	\end{assumption}
	
	Assumption \ref{ass-conditional-M-regularity}(i) and (ii) are standard regularity conditions.
	Assumption \ref{ass-conditional-M-regularity}(iii) is a local quadratic curvature condition on the population criterion, which is standard in $M$-estimation theory \citep{vandervaart1998}.
	Assumption \ref{ass-conditional-M-regularity}(iv) is an identification condition for $\vartheta_0(\bx)$.
	Assumption \ref{ass-conditional-M-regularity} is satisfied when $\ell(U, \vartheta) = (U - \vartheta)^{2}$
	and $U$ is uniformly bounded.
	It is also satisfied when $\ell(U, \vartheta) = \rho_{\tau}(U - \vartheta)$ and
	the conditional density function of $U$ given $\bX = \bx$ is bounded away from zero for $U = u$ in a neighborhood of the $\tau$-th quantile of the distribution $U\mid \bX=\bx$ for almost surely all $\bx$.
	
	\begin{theorem}
		\label{theorem-rate-conditional-M-functional}
		Suppose that Assumptions
		\ref{ass-bounded-sieves}-\ref{ass-conditional-M-regularity}
		are satisfied.
		Let $C_{\Theta}$ denote the diameter of $\Theta$.
		Then there exists a constant
		\(
		C_\vartheta>0\),
		depending only on
		\(
		(D,K_1, K_2,
		\kappa_\ell, B_\ell,V_\ell, A_\ell, \Delta_{\ell}, C_{\Theta})
		\),
		such that
		\bas
		\Pr\left( \|
		\widehat\vartheta-\vartheta_0
		\|_{L_2(P_X)}
		\le
		C_\vartheta
		\{
		a_n
		+
		\sqrt{\log(11/\varepsilon)}\,\delta_n
		\} \right) \geq 1-\varepsilon,
		\eas
		for every $\varepsilon$ satisfying $0< \varepsilon <1$ and
		\(
		\log\left(11/{\varepsilon}\right)\le n
		\).
	\end{theorem}
	
	Theorem \ref{theorem-rate-conditional-M-functional} shows that the conditional feature estimator attains the same order of non-asymptotic error as $\widehat\phi$. For a fixed $\by\in\mathcal{Y}$, setting $T(\bY) =I(\bY\leq\by)$, $\ell(u,\vartheta)=(u-\vartheta)^2$, and $\Theta=[0,1]$ in the above framework yields $\vartheta_0(\bx)= F_{Y\mid X, 0}(\by\mid\bx)$. Under the same specification, the $M$-estimator in \eqref{proposed-condiitonal-mean} reduces to $\widehat F_{Y\mid X}(\by\mid\bx)$. Consequently, the error bound for $\widehat F_{Y\mid X}(\by\mid\bx)$ follows immediately from Theorem \ref{theorem-rate-conditional-M-functional}.
	
	\begin{corollary}
		\label{corollary-rate-conditional-distribution}
		Suppose that Assumptions \ref{ass-bounded-sieves}-\ref{ass-approximation}
		are satisfied. Then there exists a constant
		\( C_d>0\),
		depending only on
		\( (D, K_1, K_2)\),
		such that
		\bas
		\inf_{\by\in\mathcal{Y}}\Pr\left(
		\|\widehat F_{Y\mid X}(\by\mid\cdot)- F_{Y\mid X,0}(\by\mid\cdot)	\|_{L_2(P_X)}
		\le
		C_d
		\{
		a_n
		+
		\sqrt{\log(11/\varepsilon)}\,\delta_n
		\} \right) \geq 1-\varepsilon,
		\eas
		for any $\varepsilon$ satisfying
		$0<\varepsilon <1$ and
		\(
		\log\left(11/{\varepsilon}\right)\le n
		\).
	\end{corollary}
	
	\subsection{Applications}
	\label{sec-application}
	
	The proposed estimation procedure is versatile enough to address many
	important problems, including estimation of the mutual density
	ratio (MDR) and mutual information (MI), and independence testing.
	The MDR, defined as the joint-to-product density ratio,
	pinpoints where and how strongly dependence occurs; MI,
	its expected log, is a cornerstone of information theory
	and statistical learning, widely used in feature selection,
	image registration, clustering, and independence testing.
	While the MDR captures fine-grained local structure,
	MI provides a single comparable scalar for optimization
	and decision-making. Together, they offer both local
	characterization and global quantification of dependence.
	
	\vspace{0.3em}
{\it Estimation of MDR.}
	The MDR of $(\bX, \bY)$ is defined as
	\(
	m_{0}(\bx, \by) = f_{0}(\bx, \by) /\{f_{x,0}(\bx) f_{y,0}(\by) \}.
	\)
	Under model \eqref{model-joint-density}, the mutual density ratio can be expressed as
	\ba
	\label{representation-mutual-density-ratio}
	m_{0}(\bx, \by)  =  \frac{ \exp\{ \phi_{0}(\bx, \by) \}\iint  f_{1,0}(\bu) f_{2,0}(\bv) \exp\{ \phi_{0}(\bu, \bv) \} d\bu d\bv}{
		\int   f_{2,0}(\bv) \exp\{ \phi_{0}(\bx, \bv) \}   d\bv \cdot   \int   f_{1,0} (\bu) \exp\{ \phi_{0}(\bu, \by) \}   d\bu  }.
	\ea
	With our MLEs $\widehat{\phi}$, $\widehat{F}_{2}$ and $\widehat{F}_{1}$,
	we propose to estimate the mutual density ratio by
	\ba
	\label{estimator-mutual-density}
	\widehat m(\bx, \by) =   \frac{ \exp\{ \widehat{\phi}(\bx, \by) \}\iint   \exp\{ \widehat{\phi}(\bu, \bv) \} d\widehat F_2(\bv) d\widehat F_1(\bu)}{
		\int   \exp\{ \widehat{\phi}(\bx, \bv) \}   d\widehat F_2(\bv) \cdot
		\int    \exp\{\widehat{\phi}(\bu, \by) \}   d\widehat F_1(\bu)  }.
	\ea
	\begin{theorem}
		\label{theorem-rate-hat-m}
		Under the conditions of Theorem \ref{prop-rate-hat-phi},
		there exists a constant $C_m>0$, depending only
		on $(D,K_1,K_2)$, such that
		\bas
		\Pr\left(
		d(\widehat m,m_0)
		\le C_m \{ a_n + \sqrt{\log(11/\varepsilon)}\,\delta_n \}
		\right) \geq
		1-\varepsilon,
		\eas
		for every $\varepsilon$ satisfying
		$0<\varepsilon<1$ and
		\( \log\left(11/\varepsilon\right)\le n\).
	\end{theorem}
	Theorem~\ref{theorem-rate-hat-m} indicates that the error bound for the proposed estimator of MDR achieves the same order of convergence rate as the estimator of interaction function.
	
		\vspace{0.3em}
	{\it Estimation of MI.}
	One prominent application of the MDR estimation is the estimation of MI \citep{shannon1948, cover2006},
	which is defined as
	$
	\operatorname*{MI} = \e[ \log\{m_{0}(\bX, \bY)\} ].
	$
	Given our estimator of MDR, a natural estimator of MI is
	$
	\widehat{\operatorname*{MI}} = (1/n)\sum_{i=1}^{n}\log\{\widehat{m}(\bX_{i}, \bY_{i})\}.
	$
	Theorem \ref{theorm-bound-mutual-information}
	implies that our MI estimator has the same order of non-asymptotic error as that of our MDR estimator.
	
	\begin{theorem}
		\label{theorm-bound-mutual-information}
		Under the conditions of Theorem \ref{prop-rate-hat-phi},
		there exists a constant $C_{\rm MI}>0$,
		depending only on $(D,K_1,K_2)$, such that
		\bas
		\Pr\left(
		|\widehat{\operatorname*{MI}} - \operatorname*{MI} |\leq C_{\rm MI}	\{
		a_n+ \sqrt { \log(12/\varepsilon)}\,\delta_n
		\} \right) \geq 1-\varepsilon,
		\eas
		for every $\varepsilon $ satisfying
		$0<\varepsilon <1$ and
		\(
		\log(12/\varepsilon)\le n\).
	\end{theorem}
	
		\vspace{0.3em}
{\it Independence testing.}
	Our estimation framework can be readily applied to independence testing.
	Specifically, we consider testing
	\(
	H_0:\bX\perp\bY
	~\text{versus}~
	H_1:\bX\not\perp\bY.
	\)
	Under \eqref{model-joint-density}, it follows immediately that
	$\phi \equiv 0$ implies $\bX\perp\bY$.
	Conversely, if $\bX$ and $\bY$ are independent, then \eqref{model-joint-density} yields
	$
	\phi(\bx, \by) = \log\{f_{x}(\bx)/f_{1}(\bx)\}+\log\{f_{y}(\by)/f_2(\by)\}+\log(Z_{\phi}),
	$
	which, together with the constraint $\phi \in \mathscr P$, forces $\phi\equiv 0$.
	Consequently, under \eqref{model-joint-density},
	testing the independence of $\bX$ and $\bY$ is equivalent to testing $\phi\equiv 0$.
	Suppose that \(0\in\mathscr F_n\subseteq\mathscr P\).
	To test the independence of $\bX$ and $\bY$, we propose the profile likelihood-ratio statistic
	\ba
	\label{definition-profile-LR-independence}
	\mathcal T_n
	=
	2
	\{
	\ell_{\mathrm{profile},2}(\widehat\phi)
	-
	\ell_{\mathrm{profile},2}(0)
	\},
	\ea
	where $\widehat{\phi}$ is defined in \eqref{defi-MLE-profile-likelihood-2}.
	Since $0\in\mathscr F_n$, $\mathcal T_n$ is nonnegative.
	Under the null hypothesis $H_0$, $\mathcal T_n$ is expected to be small;
	accordingly, $H_0$ is rejected for sufficiently large values of $\mathcal T_n$.
	Because the null distribution of $\mathcal T_n$ is analytically intractable,
	we resort to a permutation procedure to compute the $p$-value of $\mathcal T_n$.
	Algorithm \ref{alg-permutation-independence} provides a detailed description of the proposed test.
	
	\begin{theorem}
		\label{theorem-validity-permutation-independence}
		Suppose that
		\(
		(\bX_1,\bY_1),\ldots,(\bX_n,\bY_n)
		\)
		are i.i.d. and that the null hypothesis
		\(
		H_0:\bX\perp\bY
		\)
		holds. Let \(\widehat p_{\mathrm{perm}}\) be the permutation $p$-value defined in
		Algorithm \ref{alg-permutation-independence}. Then, for every $\alpha\in(0,1)$,
		\(
		\Pr
		\left(
		\widehat p_{\mathrm{perm}}
		\le\alpha
		\right)
		\le
		\alpha.
		\)
		Consequently, the permutation procedure provides a finite-sample valid
		level-$\alpha$ test of independence.
		
	\end{theorem}
	
	\begin{algorithm}[htbp]
		\caption{Permutation test for independence testing}
		\label{alg-permutation-independence}
		
		\begin{algorithmic}[1]
			
			\Require
			Sample
			$\mathcal D_n=\{(\bX_i,\bY_i):i=1,\ldots,n\}$,
			number of permutations $B$, and significance level $\alpha$.
			
			\State
			Compute the MLE $\widehat{\phi}$ and the test statistic
			\(
			\mathcal T_n\)
			
			\For{$b=1,\ldots,B$}
			
			\State
			Generate a random permutation $\pi_b$ of
			$\{1,\ldots,n\}$ and define a permuted sample as
			$
			\mathcal D_n^{\pi_b}
			=
			\{(\bX_i,\bY_{\pi_b(i)}):i=1,\ldots,n\}.
			$
			
			\State
			Using $\mathcal D_n^{\pi_b}$, compute
			$
			\widehat\phi^{\pi_b}
			\in
			\operatorname*{argmax}_{\phi\in\mathscr F_n}
			\ell_{\mathrm{profile},2}^{\pi_b}(\phi).
			$
			
			\State
			Calculate the permuted statistic
			$
			\mathcal T_n^{\pi_b}
			=
			2
			\{
			\ell_{\mathrm{profile},2}^{\pi_b}
			(\widehat\phi^{\pi_b})
			-
			\ell_{\mathrm{profile},2}^{\pi_b}(0)
			\}.
			$
			
			\EndFor
			
			\State
			Compute the permutation $p$-value
			$
			\widehat p_{\mathrm{perm}}
			=
			\frac{1}{B+1}\{
			1+
			\sum_{b=1}^{B}
			I(\mathcal T_n^{\pi_b}\ge\mathcal T_n)
			\}.
			$
			
			\If{$\widehat p_{\mathrm{perm}}\le\alpha$}
			
			\State
			Reject $H_0:\bX\perp\bY$.
			
			\Else
			
			\State
			Do not reject $H_0:\bX\perp\bY$.
			
			\EndIf
			
		\end{algorithmic}
	\end{algorithm}

	\section{Deep feedforward neural network sieves}
	\label{sec-dnn}
	
	The above results hold for general nonparametric sieve spaces satisfying
	Assumption \ref{ass-entropy-full-sieves}.
	For illustration, in this section we consider a specific example of such
	a sieve space, namely a class of feedforward neural networks
	(FNNs; \citealp{schmidtHieber2020, yan2025deep, Lang2026GCS}).
	
	Assume that $\mathcal X\times\mathcal Y=[0,1]^d$ for some $d\geq 2$.
	Let $\bz = (\bx^{\top}, \by^{\top})^{\top}$, $K\in\mathbb{N}_{+}$, and $\bt=(t_{0},\cdots, t_{K}, t_{K+1})\in\mathbb{N}_{+}^{K+2}$.
	We define $g:\mathbb{R}^{t_{0}}\mapsto\mathbb{R}^{t_{K+1}}$, a $(K+1)$-layer FNN with layer widths specified by $\bt$, as
	$g(\bz)=\bW_{K}\bg_{K}(\bz)+\bmu_{K},
	\bg_{K}(\bz)=\sigma\left(\bW_{K-1}\bg_{K-1}(\bz)+\bmu_{K-1}\right),\cdots, \bg_{1}(\bz)=\sigma\left(\bW_{0}\bz+\bmu_{0}\right).
	$
	The matrices $\bW_{j}\in\mathbb{R}^{t_{j+1}\times t_{j}}$
	and the vectors $\bmu_{j}\in\mathbb{R}^{t_{j+1}}$ for $j=0,\cdots, K$
	constitute the parameters of the FNN.
	The activation function $\sigma$ operates component-wise,
	meaning that $\bg_{j}=(g_{j1},\cdots, g_{jt_{j}})^{\top}$
	maps from $\mathbb{R}^{t_{j-1}}$ to $\mathbb{R}^{t_{j}}$
	for each $j$.
	Here we consistently set $t_{0}=d$ and $t_{K+1}=1$,
	and choose $\sigma$ to be the popular rectified linear unit
	$
	x \vee 0
	$.
	Formally, we define
	$\mathcal{F}(K,\bt, \tilde{D}) = \{ g \mid g~\text{is a}~(K+1)\text{-layer FNN}~\text{with width}~\bt~\text{such that } \|g\|_{\infty}\leq \tilde D \}$ for $K\in\mathbb{N}_{+}$, $\bt\in\mathbb{N}_{+}^{K+2}$ and $\tilde{D}>0$.
	
	For deterministic sequences $K_n$ and $\bt_n = (t_{n,0}, \cdots, t_{n,K_{n}}, t_{n, K_{n}+1})$, let $\mathscr{F}_{n}^{(0)} = \mathcal{F}(K_n,\bt_n,\tilde{D})$ in the definition of $\mathscr F_n$. Write
	\(
	\overline t_n
	=
	\max_{1\le j\le K_n}t_{n,j}.
	\)
	To mitigate the curse of dimensionality, we assume that the true $\phi_{0}$ belongs to the hierarchical composition function (HCF) classes \citep{yan2025deep}.
	
	\begin{definition}[H$\rm \ddot{o}$lder classes]
		Let $a$ and $M$ be positive constants, and let $\mathbb{D}\subset\mathbb{R}^{d}$.
		A H$\rm \ddot{o}$lder class of smooth functions is defined as
		\bas
		\mathcal{H}_{d}^{a}(\mathbb{D}, M)
		=\left\{g:\mathbb{D}\mapsto\mathbb{R}\bigg|\sum\limits_{\blamb:|\blamb|<a}\|\partial^{\blamb}g\|_{\infty}
		+\sum\limits_{\blamb:|\blamb|=\lfloor a \rfloor}\sup_{\bz_{1},\bz_{2}\in\mathbb{D},\bz_{1}\neq \bz_{2}}\frac{|\partial^{\blamb}g(\bz_{1})-\partial^{\blamb}g(\bz_{2})|}{\|\bz_{1}-\bz_{2}\|_{\infty}^{a-\lfloor{a}\rfloor}}\leq M\right\},
		\eas
		where
		$\blamb=(\lambda_{1},\cdots,\lambda_{d})$ with $\lambda_i \in \mathbb{N}$,
		$|\blamb|=\sum_{i=1}^{d}\lambda_{i}$, $\partial^{\blamb} := \partial^{\lambda_{1}}\cdots\partial^{\lambda_{d}}$, and $\lfloor a \rfloor$ denotes the largest integer strictly smaller than $a$.
	\end{definition}
	
	If a conditional mean regression function belongs to
	$\mathcal{H}_{d}^{a}(\mathbb{D}, M)$,
	the optimal minimax rate of convergence
	for estimating it is $n_P^{-\frac{a}{2a+d}}$ \citep{stone1982optimal}. When $d$ is large, this rate can be extremely slow.
	To mitigate this curse of dimensionality,
	we consider HCF classes.
	
	\begin{definition}[HCF class, \citealp{yan2025deep}]
		Let $q\in\mathbb{N}$, $M>0$, and let $\Sigma$ be a $q$-height tree with nodes belonging to $(0, \infty)\times\mathbb{N}$. Then a HCF class $\mathcal{G}(q, \Sigma, M)$ is recursively defined as follows. For $q=1$ and $\Sigma = ((\theta, k))$, define
		$
		\mathcal{G}(q, \Sigma, M) = \{\phi:[0,1]^{d}\mapsto \mathbb{R}\Big| \phi(\bz) = g(z_{\varrho(1)}, \cdots, z_{\varrho(k)}), ~ g\in\mathcal{H}_{k}^{\theta}([0,1]^{k}, M), \varrho:[k]\mapsto[d] \},
		$
		where $z_{\varrho(j)}$ is the $\varrho(j)$-th component of $\bz$.
		For $q\geq 2$, given $\Sigma = ((\theta, k); \Sigma_{1}, \cdots, \Sigma_{k})$ with $(\theta, k)\in (0, \infty)\times \mathbb{N}$, define
		$
		\mathcal{G}(q, \Sigma, M) = \{h:[0,1]^{d}\mapsto \mathbb{R}: h(\bz) = g(g_{1}(\bz), g_{2}(\bz), \cdots, g_{k}(\bz)), ~\text{where} ~g\in\mathcal{H}_{k}^{\theta}(\mathbb{R}^{k}, M), ~\text{and}~ g_{j}\in\mathcal{G}(q_{j}, \Sigma_{j}, M)~\text{with}~q_{j}\leq q-1 ~\text{for} ~1\leq j\leq k \}.
		$
	\end{definition}
	
	For a HCF class $\mathcal{G}(q, \Sigma, M)$, for each node $G = (\theta_{G}, k_{G})$ of $\Sigma$, define the effective smoothness index as $\widetilde{\theta}_{G} = \theta_{G}\times\prod_{(\theta, k)\in\mathcal{B}(G)}(1\wedge \theta)$, where $\mathcal{B}(G)$ denotes the ancestor set of $G$ in the tree $\Sigma$.
	We define the intrinsic smoothness-dimension ratio as
	\ba
	\label{intrinsic-dim-ratio}
	\gamma = \min_{G\in \text{node}(\Sigma)} \frac{\widetilde{\theta}_{G}}{ k_{G}} = \frac{\widetilde{\theta}_{G^{\star}}}{ k_{G^{\star}}},
	\ea
	where $\text{node}(\Sigma)$ denotes the set of all nodes of $\Sigma$
	and $G^{\star}\in\operatorname*{argmin}_{G\in \text{node}(\Sigma)} \frac{\widetilde{\theta}_{G}}{ k_{G} }$.
	To some extent, $k_{G^{\star}}$ characterizes the intrinsic dimension
	of $\mathcal{G}(q,\Sigma, M)$.
	Let
	\(
	r_n
	=
	n^{-\frac{\gamma}{ 2\gamma+1 }}
	(\log n)^{ \frac{6\gamma}{ 2\gamma+1}}.
	\)
	
	\begin{assumption}
		\label{ass-dnn-hcf-architecture}
		(i)
		\(
		\phi_0\in
		\mathcal G(q,\Sigma,M)\cap\mathscr P.
		\)
		(ii) $\tilde{D}\ge M$. (iii)
		There exist integer sequences
		$\widetilde K_n,\widetilde t_n\ge3$ such that
		\(
		K_n
		\asymp
		\widetilde K_n\log(\widetilde K_n),\)
		and \( \overline t_n
		\asymp
		\widetilde t_n\log(\widetilde t_n),
		\)
		and
		\(
		\widetilde K_n\widetilde t_n
		\asymp
		n^{1/(4\gamma+2)}
		(\log n)^{-3/(2\gamma+1)}.
		\)
	\end{assumption}
	
	Assumption \ref{ass-dnn-hcf-architecture}
	controls the approximation and stochastic errors of the FNN sieves.
	
	\begin{corollary}
		\label{corollary-rate-hat-m-dnn}
		Suppose that Assumption \ref{ass-dnn-hcf-architecture} and condition (ii) of Assumption
		\ref{ass-true-function-identification} are satisfied. Let $\mathscr{F}_{n}^{(0)} = \mathcal{F}(K_n,\bt_n,\tilde{D})$, and let $n$ be larger than a fixed integer.
		Then there exists a constant
		\(
		C_{\mathrm{NN}, 1}>0\),
		depending only on
		\(
		(\tilde{D},\ K_1,\ K_2)
		\), such that
		\[
		\Pr\left( d(\widehat m,m_0)
		\le
		C_{\mathrm{NN}, 1}
		\sqrt{\log(11/\varepsilon)}
		\cdot r_n \right) \geq 1-\varepsilon,
		\]
		for every $\varepsilon$ satisfying
		$0<\varepsilon<1$ and
		\(
		\log(11/\varepsilon)\le n\).
		
	\end{corollary}
	
	Corollary~\ref{corollary-rate-hat-m-dnn} shows that, by balancing the stochastic and approximation errors, the proposed MDR estimator attains the convergence rate $r_{n}$, which coincides, up to logarithmic factors, with the minimax rate for regression over the HCF class \citep{schmidtHieber2020} and does not depend explicitly on the ambient covariate dimension $d$. Consequently, our result mitigates the curse of dimensionality when the true function $\phi_{0}$ exhibits an underlying low-dimensional structure \citep{schmidtHieber2020, Zhong2022, yan2025deep, Lang2026GCS}. Similarly, the non-asymptotic error bounds for the MI and the conditional feature estimators under FNN sieves are stated as follows.
	
	\begin{corollary}
		\label{corollary-rate-hat-mi-dnn}
		Under the assumptions of Corollary \ref{corollary-rate-hat-m-dnn},
		there exists a constant
		\(
		C_{\mathrm{NN}, 2}>0\),
		depending only on
		\(
		(\tilde{D},\ K_1,\ K_2)
		\), such that
		\[
		\Pr\left(
		\left|\widehat{\operatorname*{MI}} - \operatorname*{MI}\right|
		\le
		C_{\mathrm{NN}, 2}
		\sqrt{\log(12/\varepsilon)}
		\cdot r_n \right) \geq 1-\varepsilon,
		\]
		for every $\varepsilon$ satisfying $0<\varepsilon <1$ and
		\(
		\log(12/\varepsilon)\le n\).
	\end{corollary}
	
	\begin{corollary}
		\label{corollary-rate-condition}
		Under Assumptions \ref{ass-conditional-M-regularity} and
		\ref{ass-dnn-hcf-architecture}, condition (ii) of Assumption
		\ref{ass-true-function-identification}, and
		$\mathscr{F}_{n}^{(0)} = \mathcal{F}(K_n,\bt_n,\tilde{D})$
		with $n$ larger than a fixed integer,
		there exists a constant
		\(
		C_{\mathrm{NN}, 3}>0\),
		depending only on
		\(
		(\tilde{D}, K_1, K_2,
		\kappa_\ell, B_\ell, V_\ell, A_\ell, \Delta_{\ell}, C_{\Theta})
		\),
		such that
		\[
		\Pr\left(
		\left\|
		\widehat\vartheta-\vartheta_0
		\right\|_{L_2(P_X)}
		\le
		C_{\mathrm{NN}, 3}
		\sqrt{\log(11/\varepsilon)}
		\cdot
		r_n
		\right)
		\geq
		1-\varepsilon,
		\]
		for every $\varepsilon$ satisfying $0<\varepsilon<1$ and
		\(
		\log(11/\varepsilon)\le n
		\).
		
	\end{corollary}
	
	Corollary \ref{corollary-rate-condition} shows that the proposed
	conditional feature estimators attain the desired convergence rates when
	$\phi_0$ belongs to the HCF regularity framework. Unlike existing results
	on deep neural-network estimation of conditional means or quantiles, our
	result does not require the true conditional mean and conditional
	quantile functions to satisfy separate regularity assumptions. Instead,
	a single structural assumption on $\phi_0$ allows both conditional
	features to be estimated simultaneously at favorable convergence rates.
	
	A similar result holds for the proposed conditional distribution estimator
	$\widehat F_{Y\mid X}(\by\mid\cdot)$.
	
	\begin{corollary}
		\label{corollary-rate-condition-distribution}
		
		Under Assumption \ref{ass-dnn-hcf-architecture}, condition (ii) of
		Assumption \ref{ass-true-function-identification}, and
		$\mathscr{F}_{n}^{(0)} = \mathcal{F}(K_n,\bt_n,\tilde{D})$
		with $n$ larger than a fixed integer,
		there exists a constant
		\(
		C_{\mathrm{NN}, 4}>0\),
		depending only on
		\(
		(\tilde{D},\ K_1,\ K_2)
		\),
		such that
		\[
		\inf_{\by\in\mathcal{Y}}\Pr\left(
		\left\|
		\widehat F_{Y\mid X}(\by\mid\cdot)
		-
		F_{Y\mid X,0}(\by\mid\cdot)
		\right\|_{L_2(P_X)}
		\le
		C_{\mathrm{NN}, 4}
		\sqrt{\log(11/\varepsilon)}
		\cdot r_n
		\right)
		\geq
		1-\varepsilon,
		\]
		for every $\varepsilon$ satisfying $0<\varepsilon<1$ and
		\(
		\log(11/\varepsilon)\le n
		\).
		
	\end{corollary}

	\section{Simulation study}
	\label{section-simu}
	
	In this section, we evaluate the numerical performance of the proposed methods for estimating the conditional features and other targets considered in Section \ref{sec-application}, under neural network sieves. For the implementation of the proposed profile likelihood estimators \eqref{defi-MLE-profile-likelihood-2} under neural network sieves, we refer the reader to Algorithm 1 in the Supplementary Material. To ensure fair comparisons, our proposed method and all competing methods based on neural networks share the same network architecture, hyperparameter settings, and Adam training parameters across all scenarios. Additional details are provided in the Supplementary Material.
	
	\subsection{Estimation performance}
	\label{sec-simulation-conditional-mean}
	
	Throughout this subsection, we set the training sample size to \(1000\) and assess the relative RMSEs of all estimators under comparison on an independent test sample of size \(2000\). Each RMSE is computed from \(200\) simulation replicates. A detailed description of the relative RMSEs and other evaluation metrics is provided in the Supplementary Material. To ensure fair comparisons, all methods use the same training-to-validation sample size ratio and are evaluated at the same test covariates within each replication.
	Let \(\operatorname{TN}(-2.5, 2.5)\) denote the standard normal distribution truncated to \([-2.5,2.5]\), \(\bX=(X_1,\ldots,X_p)^\top\), and \(\bI_p\) the \(p\times p\) identity matrix.
	
		\vspace{0.3em}
{\it Scalar Response.}
	We first investigate the estimation of conditional characteristics for a scalar \(Y\).
	Our estimator \eqref{proposed-condiitonal-mean} (Proposed) is compared with
	the logistic-loss (Logistic) and least-squares-loss MDR (Least-squares) estimators of \cite{Han2025},
	as well as the LinCDE boosting estimator (LinCDE) of \cite{GaoHastie2022}. All of
	these procedures recover the conditional characteristics from an estimated
	conditional distribution or density. The two estimators of \cite{Han2025}
	first fit the MDR and then combine it with an estimate of the marginal
	density of \(Y\) to construct a normalized conditional density, whereas
	LinCDE estimates the conditional density directly.
	
	\begin{example} \label{uni-response}
		Let \(X_1, \ldots, X_p\) and \(\epsilon\) be i.i.d. as \(\operatorname{TN}(-2.5, 2.5)\). We consider the following three models: (M1) \(Y = X_1^2 + \epsilon\) with \(p = 1\); (M2) \(Y = X_1^2 + \exp\{(X_2 + X_3)/3\} + \sin(X_4 + X_5) + \epsilon\) with \(p = 5\); and (M3) \(Y = \bigl(\sum_{j=1}^5 X_j / \sqrt{5}\bigr)^2 + \epsilon\) with \(p = 5\).
	\end{example}
	
	Models M1 and M2 are models (M1) and (M2) in the simulation studies of \cite{Han2025}.
	For data generated from Example \ref{uni-response}, Table~\ref{tab:conditional_characteristics_rrmse} reports the simulated relative RMSEs of the conditional mean and conditional quantile estimators.
	The proposed method attains the smallest relative RMSE in every scenario considered, demonstrating consistently superior performance over the competing methods. The improvement is particularly pronounced under Models M1 and M3. Under M1, relative to the best-performing competitor in each scenario, the proposed method reduces the relative RMSE by at least approximately \(20.8\%\) across all  estimands; the corresponding reduction under M3 is at least approximately \(9.9\%\). These results show that our method delivers substantial and stable gains in estimation accuracy for both conditional means and conditional quantiles.
	
	\begin{table}[htbp]
		\centering
		\caption{Simulated relative RMSEs and their Monte Carlo standard errors (in parentheses) for conditional mean and conditional quantile estimation under Models M1, M2, and M3. All numbers are reported in units of 0.01.}
		\label{tab:conditional_characteristics_rrmse}
		\setlength{\tabcolsep}{10pt}
		\begin{tabular}{ cccccc}
			\toprule
			Model & Estimand & Proposed & Logistic & Least-squares & LinCDE \\
			\midrule
			\multirow{3}{*}{M1}
			& Mean
			& ${\bf 8.58}(0.16)$
			& ${12.46}(0.23)$
			& ${11.81}(0.30)$
			& ${31.20}(0.13)$ \\

			& Quantile$_{0.5}$
			& ${\bf 9.05}(0.15)$
			& ${12.18}(0.20)$
			& ${11.43}(0.23)$
			& ${31.56}(0.13)$ \\

			& Quantile$_{0.75}$
			& ${\bf 7.12}(0.11)$
			& ${ 9.54}(0.16)$
			& ${ 9.08}(0.20)$
			& ${24.05}(0.11)$ \\
			
			\hline
			
			\multirow{3}{*}{M2}
			& Mean
			& ${\bf 19.50}(0.11)$
			& ${20.41}(0.13)$
			& ${22.40}(0.26)$
			& ${28.43}(0.08)$ \\

			& Quantile$_{0.5}$
			& ${\bf 20.42}(0.10)$
			& ${20.75}(0.13)$
			& ${23.27}(0.23)$
			& ${28.43}(0.08)$ \\

			& Quantile$_{0.75}$
			& ${\bf  15.93}(0.07)$
			& ${16.36}(0.09)$
			& ${19.37}(0.22)$
			& ${23.99}(0.07)$ \\
			
			\hline
			
			\multirow{3}{*}{M3}
			& Mean
			& ${\bf 27.09}(0.29)$
			& ${32.29}(0.27)$
			& ${30.19}(0.49)$
			& ${70.24}(0.24)$ \\

			& Quantile$_{0.5}$
			& ${\bf 28.99}(0.30)$
			& ${33.03}(0.26)$
			& ${32.18}(0.47)$
			& ${71.90}(0.26)$ \\

			& Quantile$_{0.75}$
			& ${\bf  19.47}(0.21)$
			& ${23.50}(0.19)$
			& ${24.48}(0.36)$
			& ${53.66}(0.24)$ \\
			
			\bottomrule
		\end{tabular}
	\end{table}
	
		\vspace{0.3em}
{\it Multivariate Response.}
	We next investigate the estimation of conditional characteristics for a multivariate response \(\bY\). Our method (Proposed) is compared with nearest-neighbor kernel conditional density estimation \cite[NNKCDE]{dalmasso2020conditional} and distributional random forests \cite[DRF]{cevid2022distributional}. For NNKCDE, we implement the estimator following its original formulation and select the number of nearest neighbors and the kernel bandwidth by minimizing the conditional density estimation loss on an independent validation sample, as recommended in \cite{dalmasso2020conditional}. For DRF, we use the R package \texttt{drf}, with its main tuning parameters fixed at the recommended settings of \cite{cevid2022distributional}. The estimated conditional distributions are then used to obtain the conditional characteristics of interest. Since the competing methods considered in the scalar scenario are designed only for scalar responses, they are not included in this simulation.
	
	\begin{example}\label{multi-response}
		Let the components \(X_1,\ldots,X_p\) of \(\bX\) be i.i.d. as \(\operatorname{TN}(-2.5, 2.5)\). Define
		\(
		V={\bX^\top\bw}/{(\sqrt{v}\|\bw\|_2)},
		\)
		where \(v\) denotes the variance of \(\operatorname{TN}(-2.5, 2.5)\) and \(\bw\) is a prespecified index vector. The response \(\bY=(Y_1,Y_2,Y_3,Y_4)^\top\) is generated according to
		\bas
		Y_i &=& V^2+ U_0\cdot \{ 2+1.5\tanh(V) \}+U_1+\varepsilon_i, \quad i=1, 2, \\
		Y_j &=& V^2+ U_0 \cdot \{2+1.5\sin(V) \}+U_2+\varepsilon_j, \quad j = 3,4,
		\eas
		where \(U_0,U_1,U_2\), and \(\varepsilon_1,\ldots,\varepsilon_4\) are i.i.d. as \(\operatorname{TN}(-2.5, 2.5)\). We consider \(p\in\{3,10 \}\) and two choices of the weight vector \(\bw\): \((1,\ldots,1)^\top\) (Equal) and \((1,\ldots, p)^\top\) (LinearRamp).
	\end{example}
	
	We generate data from Example \ref{multi-response} and focus on the estimation of the conditional covariance functions
	\(
	\operatorname*{Cov}_{i,j}(\bX)
	:=
	\operatorname*{Cov}(Y_i,Y_j\mid\bX)
	\)
	for
	\(
	(i,j)\in\{(1,2),(2,3),(3,4)\}.
	\)
	Table~\ref{tab:conditional_covariance_rrmse} presents the simulated relative RMSEs of all estimators.
	As in the univariate case, the proposed estimator again achieves the smallest relative RMSE in every scenario, consistently outperforming NNKCDE and DRF. Relative to the best-performing competitor (NNKCDE throughout the table), it reduces the relative RMSE by at least approximately \(13.25\%\) across all scenarios, with reductions ranging from
	\(14.82\%\) to \(17.44\%\) at \(p=10\). The gains over DRF are substantially larger, ranging from approximately \(27.78\%\) to \(36.9\%\). Although all methods show increasing relative RMSEs as the covariate dimension grows, the proposed method retains the best performance throughout. These results indicate that the proposed method effectively recovers conditional characteristics for multivariate responses.
	
	\begin{table}[htbp]
		\centering
		\caption{Simulated relative RMSEs, with Monte Carlo standard errors in parentheses, for conditional covariance estimation under different index structures and covariate dimensions. All numbers are reported in units of 0.01.}
		\label{tab:conditional_covariance_rrmse}
		\setlength{\tabcolsep}{10pt}
		\begin{tabular}{cccccc}
			\toprule
			Index& Dimension & Estimand & Proposed & NNKCDE & DRF \\ \midrule
			
			\multirow{6}{*}{Equal}
			& \multirow{3}{*}{3} & $\operatorname*{Cov}_{1,2}(\bX)$ & ${\bf 21.86}(0.31)$ & ${25.20}(0.20)$ & ${34.56}(0.35)$ \\
			& & $\operatorname*{Cov}_{2,3}(\bX)$ & ${\bf 24.69}(0.35)$ & ${29.05}(0.21)$ & ${39.13}(0.38)$ \\
			& & $\operatorname*{Cov}_{3,4}(\bX)$ & ${\bf 23.19}(0.30)$ & ${27.28}(0.18)$ & ${35.61}(0.32)$ \\
			
			\cmidrule(lr){2-6}
			& \multirow{3}{*}{10}& $\operatorname*{Cov}_{1,2}(\bX)$ & ${\bf 34.01}(0.31)$ & ${40.38}(0.22)$ & ${51.55}(0.29)$ \\
			& & $\operatorname*{Cov}_{2,3}(\bX)$ & ${\bf 38.51}(0.35)$ & ${46.59}(0.25)$ & ${59.12}(0.32)$ \\
			& & $\operatorname*{Cov}_{3,4}(\bX)$ & ${\bf 35.07}(0.33)$ & ${42.48}(0.22)$ & ${53.23}(0.31)$ \\
			
			\midrule
			
			\multirow{6}{*}{LinearRamp}
			& \multirow{3}{*}{3} & $\operatorname*{Cov}_{1,2}(\bX)$ & ${\bf 21.50}(0.31)$ & ${25.13}(0.20)$ & ${30.40}(0.36)$ \\
			& & $\operatorname*{Cov}_{2,3}(\bX)$ & ${\bf 24.17}(0.36)$ & ${28.96}(0.21)$ & ${34.37}(0.38)$ \\
			& & $\operatorname*{Cov}_{3,4}(\bX)$ & ${\bf 22.67}(0.33)$ & ${27.05}(0.18)$ & ${31.39}(0.34)$ \\
			
			\cmidrule(lr){2-6}
			& \multirow{3}{*}{10}& $\operatorname*{Cov}_{1,2}(\bX)$ & ${\bf 34.21}(0.30)$ & ${40.16}(0.21)$ & ${49.33}(0.30)$ \\
			& & $\operatorname*{Cov}_{2,3}(\bX)$ & ${\bf 38.79}(0.35)$ & ${46.45}(0.24)$ & ${56.66}(0.33)$ \\
			& & $\operatorname*{Cov}_{3,4}(\bX)$ & ${\bf 35.33}(0.32)$ & ${42.37}(0.21)$ & ${51.12}(0.32)$ \\
			
			\bottomrule
		\end{tabular}
	\end{table}
	
		\vspace{0.3em}
{\it Estimation of MDR and MI.}
	We now evaluate the proposed method for estimating the MDR and the MI.
	The proposed estimator is compared with three competing procedures: the
	Bregman-divergence deep neural network estimators of \cite{Han2025} under
	two choices of divergence, namely the logistic loss and the least-squares
	loss. Note that, with the least-squares loss, the method of \cite{Han2025}
	reduces to the density-ratio fitting method of \cite{Tsai2020}. A kernel
	plug-in estimator (Plug-in), which separately estimates the joint and marginal densities
	using Gaussian kernels, is also included in the comparison. Each estimated MDR
	is then used to construct the corresponding MI estimator via direct  plug-in.
	
	\begin{example}\label{ex-mdr-mi}
		Consider a correlated Gaussian model. For a given dimension \(p\) and
		dependence parameter \(\rho\),
		let \((\bY^\T, \bX^\T)^\T\) follow a multivariate normal distribution
		with zero mean, \(\cov(\bY) = \cov(\bX) = \bI_p\), and \(\cov(\bY, \bX) = \rho \bI_p\).
		Under this model, the true log-MDR is
		\[
		\begin{aligned}
			\log m_0(\bx,\by)
			=
			-\frac{p}{2}\log(1-\rho^2)
			+
			\frac{
				\rho\,\bx^{\T}\by
				-\rho^2\{\|\bx\|^2+\|\by\|^2\}/2
			}{
				1-\rho^2
			},
		\end{aligned}
		\]
		and the MI is
		\(
		-p\log(1-\rho^2)/2.
		\)
		We consider \(\rho\in\{0.5,0.75\}\) and \(p\in\{1,2,5,9\}\).
	\end{example}
	
	Table \ref{tab:simulation_results_mdr_mi} reports the simulation results for MDR and MI estimation based on data generated from Example \ref{ex-mdr-mi}. The proposed MDR estimator achieves the lowest relative RMSE in all scenarios, reducing the error by approximately \(7.6\%\) to \(40.8\%\) relative to the best-performing competitor, with particularly pronounced improvements when \(\rho=0.75\). For MI estimation, our method achieves the lowest relative RMSE in five of the eight scenarios, while the Logistic method is the best in the remaining three. These results demonstrate the superior performance of the proposed method for both MDR and MI estimation.
	
	\begin{table}[htbp]
		\centering
		\caption{Simulated relative RMSEs for MDR and MI estimation under each competing method. Values in parentheses are Monte Carlo standard errors for MDR estimation and raw RMSEs for MI estimation. All numbers are reported in units of 0.01.}
		\label{tab:simulation_results_mdr_mi}
		\small\setlength{\tabcolsep}{3.5pt}
		\begin{tabular}{cc cccc cccc}
			\toprule
			& & \multicolumn{4}{c}{MDR} & \multicolumn{4}{c}{MI} \\
			\cmidrule(lr){3-6} \cmidrule(lr){7-10}
			$\rho$ & $p$
			& Proposed & Logistic & Least-squares & Plug-in
			& Proposed & Logistic & Least-squares & Plug-in \\
			\midrule
			
			\multirow{4}{*}{0.5}
			& 1 & $\underset{(0.47)}{\bf 31.62}$ & $\underset{(0.38)}{34.22}$ &$\underset{(0.66)}{35.43}$ & $\underset{(0.61)}{34.35}$ & $\underset{(2.28)}{\bf 15.84}$ & $\underset{(4.20)}{29.21}$ & $\underset{(6.79)}{47.20}$ & $\underset{(3.11)}{21.59}$ \\
			& 2 & $\underset{(0.28)}{\bf 35.72}$ & $\underset{(0.26)}{39.21}$ & $\underset{(0.68)}{44.75}$ & $\underset{(0.38)}{59.26}$ & $\underset{(5.11)}{\bf 17.75}$ & $\underset{(5.49)}{19.07}$ & $\underset{(10.56)}{36.71}$ & $\underset{(28.43)}{98.84}$ \\
			& 5 & $\underset{(0.15)}{\bf 36.72}$ & $\underset{(0.21)}{43.11}$ & $\underset{(1.65)}{62.90}$ & $\underset{(0.24)}{97.04}$ & $\underset{(13.03)}{18.12}$ & $\underset{(10.67)}{\bf 14.84}$ & $\underset{(35.13)}{48.84}$ & $\underset{(244.37)}{339.78}$ \\
			& 9 & $\underset{(0.12)}{\bf 41.31}$ & $\underset{(0.23)}{51.79}$ & $\underset{(2.10)}{84.12}$ & $\underset{(0.19)}{107.47}$ & $\underset{(30.84)}{23.82}$ & $\underset{(14.90)}{\bf 11.51}$ & $\underset{(73.29)}{56.61}$ & $\underset{(278.29)}{214.96}$ \\
			
			\midrule
			
			\multirow{4}{*}{0.75}
			& 1 & $\underset{(0.24)}{\bf 19.15}$ & $\underset{(0.25)}{ 26.14}$ & $\underset{(0.32)}{ 24.15}$ & $\underset{(0.39)}{ 22.09}$ & $\underset{(2.89)}{\bf 6.99}$ & $\underset{(4.37)}{ 10.56}$ & $\underset{(6.48)}{ 15.68}$ & $\underset{(3.49)}{8.45}$ \\
			& 2 & $\underset{(0.16)}{\bf 24.19}$ & $\underset{(0.16)}{ 33.27}$ & $\underset{(0.56)}{ 40.25}$ & $\underset{(0.23)}{ 34.56}$ & $\underset{(7.77)}{ 9.40}$ &$\underset{(7.75)}{\bf 9.38}$ & $\underset{(20.69)}{ 25.03}$ & $\underset{(28.42)}{ 34.38}$ \\
			& 5 & $\underset{(0.08)}{\bf 24.08}$ & $\underset{(0.14)}{ 34.83}$ & $\underset{(0.97)}{ 51.99}$ & $\underset{(0.10)}{ 49.17}$ & $\underset{(12.82)}{\bf 6.20}$ & $\underset{(20.56)}{ 9.95}$ & $\underset{(68.76)}{ 33.27}$ & $\underset{(244.48)}{ 118.29}$ \\
			& 9 & $\underset{(0.08)}{\bf 26.89}$ & $\underset{(0.11)}{ 45.39}$ & $\underset{(0.49)}{ 52.84}$ & $\underset{(0.08)}{ 49.48}$ & $\underset{(16.94)}{\bf 4.55}$ & $\underset{(95.17)}{25.58}$ & $\underset{(110.00)}{ 29.57}$ & $\underset{(279.29)}{ 75.08}$ \\
			
			\bottomrule
		\end{tabular}
	\end{table}
	
	\subsection{Independence testing}
	\label{sec-simulation-independence}
	
	Finally, we evaluate our permutation test (Proposed) for testing
	\(
	H_0:\bX\perp\bY
	\)
	against
	\(H_1:\bX\not\perp\bY.
	\)
	We compare it with \cite{Han2025}'s permutation test (MdrMI) based
	on their deep MDR-based MI estimator,
	the distance covariance-based independence test \citep[DC]{szekely2007},
	and the Hilbert--Schmidt independence criterion \citep[HSIC]{gretton2005, gretton2008}.
	The null distributions of both Proposed and MdrMI are obtained by independently permuting the \(\bY\)-observations while leaving the \(\bX\)-observations unchanged.
	Both methods use the same \(B=100\) random permutations within each Monte Carlo replication.
	Let \(\bY\), \(\bX\), \(\bU\), and \(\bO\) be \(p\)-dimensional vectors, and let \(\bO\) follow a \(p\)-dimensional standard normal distribution and be independent of \(\bX\) and \(\bU\).
	The significance level is set to 0.05.
	
	\begin{example}
		\label{ex-it}
		Let \(p=5\) and \((\bU^\T, \bX^\T)^\T\) follow a multivariate normal distribution with zero mean, \(\cov(\bU) = \cov(\bX) = \bI_p\), and \(\cov(\bU, \bX) = \rho \bJ_p\), where \(\bJ_p\) is a \(p\times p\) matrix with all entries being 1. We consider six models:
		(M1) \(\bY = \bU\) with \(\rho = 0\);
		(M2) \(\bY = \bU\) with \(\rho = 0.1\);
		(M3) \(\bY = \bX \cdot \bO\);
		(M4) \(\bY = \log(\bX^2)\);
		(M5) \(\bY = \sin(3\bX) + 0.5\bO\);
		(M6) \(\bY = \operatorname{sgn}(\sin(3\bX)) + 0.5\bO\).
		All operations are performed elementwise.
	\end{example}
	
	Models (M1)--(M4) in Example \ref{ex-it} coincide with the four settings considered in Section S.8.4 of the supplementary material of \cite{Han2025}. The remaining two models, (M5) and (M6), are new and are designed to capture oscillatory dependence. We set the sample size to 200 or 500 and the number of simulation repetitions to 200. Table \ref{tab:simulation_results_it} reports the simulated rejection rates of the four independence tests under comparison for data generated from Example \ref{ex-it}.
	
	Under model (M1), \(\bY\) and \(\bX\) are independent, so \(H_0\) holds and the corresponding rejection rates are type I errors. At the 0.05 significance level, all four tests exhibit empirical type I error rates close to or below the nominal level, with no substantial inflation. Under the remaining five models, \(H_0\) no longer holds and the rejection rates quantify power. Under model (M2), all tests attain nearly identical powers, each close to 1. Models (M3) and (M4) are, respectively, a heterogeneous model and a nonlinear model; in these settings the Proposed test is as powerful as DC and HSIC and slightly more powerful than MdrMI.
	
	The oscillatory models (M5) and (M6) are the key discriminating cases. At \(n=200\), the methods are largely comparable, with the FNN methods ahead by a non-significant margin of about 0.10. At \(n=500\), however, a clear and significant separation emerges: both FNN methods substantially outperform HSIC and DC, in the consistent order Proposed \(\succ\) MdrMI \(\succ\) DC \(\succ\) HSIC, where \(A \succ B\) means $A$ outperforms $B$. For instance, under (M5) the rejection rates are 0.96 (Proposed), 0.92 (MdrMI), 0.76 (DC), and 0.68 (HSIC); the gaps between the FNN and classical tests, roughly \(0.20\)--\(0.28\), are several Monte Carlo standard errors. These results demonstrate that, once the sample size is sufficient for the network to resolve the oscillation, the deep-learning-based tests are markedly more sensitive to oscillatory dependence than fixed-bandwidth kernel and distance-covariance tests. Overall, the proposed test not only matches the established kernel and distance-based competitors in standard settings but also offers a distinct advantage precisely where classical fixed-bandwidth methods struggle, e.g.,  oscillatory dependence, thereby broadening the class of detectable alternatives.
	
	\begin{table}[htbp]
		\centering
		\caption{Simulated rejection rates of the Proposed, MdrMI, DC and HSIC tests when data were generated from Example \ref{ex-it}}
		\label{tab:simulation_results_it}
		\setlength{\tabcolsep}{9pt}
		\centering
		\vspace{1ex}
		\begin{tabular}{@{}c|cccc|cccc@{}}\toprule
			& \multicolumn{4}{c|}{$n=200$} & \multicolumn{4}{c}{$n=500$} \\
			Model & Proposed & MdrMI & DC & HSIC& Proposed & MdrMI & DC & HSIC\\\midrule
			M1 & 0.02& 0.04& 0.01& 0.04& 0.05& 0.06& 0.06& 0.05\\
			M2 & 1.00& 0.99& 1.00& 0.98& 1.00& 1.00& 1.00& 1.00\\
			M3 & 0.99& 0.44& 0.89& 1.00& 1.00& 0.85& 1.00& 1.00\\
			M4 & 0.92& 0.85& 1.00& 1.00& 1.00& 1.00& 1.00& 1.00\\
			M5 & 0.32& 0.35& 0.25& 0.22& 0.96& 0.92& 0.76& 0.68\\
			M6 & 0.31& 0.29& 0.29& 0.26& 0.96& 0.92& 0.76& 0.65\\
			\bottomrule
		\end{tabular}
	\end{table}
	
	\subsection{High-dimensional independence testing and the role of cross-fitting}
	\label{sec-simulation-independence-high}
	
	The rate theory of Section~\ref{sec-dnn} suggests that the proposed test should adapt to the intrinsic structure of the interaction function rather than to the ambient dimension. To examine this empirically, we consider the high-dimensional ``fair alternative'' of \cite{Ramdas2015}, also used by \cite{Han2025}, in which the dependence is confined to a single coordinate pair embedded among \(p-1\) pairs of pure noise, with \(n=1000\) and \(p\in\{100,200\}\). The full design and results, including the size tables and a learning-rate sensitivity analysis, are reported in the Supplementary Material.
	
	In this regime, how the profile likelihood-ratio statistic is evaluated becomes critical. Evaluating it in-sample---on the same data used to fit the interaction function---yields an over-optimistic statistic that is sensitive to the optimization learning rate and that \emph{loses} power as the dimension grows: for instance, at \(a=0.6\) the in-sample proposed test drops from a rejection rate of \(0.60\) at \(p=100\) to \(0.20\) at \(p=200\), and it collapses further under a larger learning rate. We therefore evaluate the statistic by \(2\)-fold cross-fitting, fitting the interaction function on one fold and evaluating the statistic on the held-out fold, averaging the two fold statistics; the same device is applied identically to the competing FNN test for a fair comparison. Cross-fitting restores both power and stability. The cross-fitted proposed test is essentially invariant to the learning rate and, crucially, sustains its power as the dimension increases (rejection rates \(0.78\)--\(0.96\) at \(p=200\), \(a=0.6\)), whereas DC, HSIC, and MdrMI test all decline sharply (to \(0.31\), \(0.18\), and \(0.08\), respectively). Thus, in high dimensions, the cross-fitted proposed test is the only procedure considered whose power does not deteriorate with dimension while retaining valid size, in line with the dimension-adaptive theory of Section~\ref{sec-dnn}.

	\section{Eye screening data analysis}
	\label{sec-real-data}
	
	We analyze the \textit{Eye Screening Data for First Year University Students in Ghana}, a cross-sectional dataset collected during university entry-level eye-health screening in 2024 and 2025 and publicly available from
	\url{https://data.mendeley.com/datasets/mfv6sb5wyc/5}. The original dataset contains 2,493 participants who completed both a questionnaire and a standardized clinical eye examination, including measurements of visual acuity, accommodative function, autorefraction, and intraocular pressure. For our analysis, the covariate vector is $\bX=(\mathrm{Age},\mathrm{AOA_R},\mathrm{AOA_L},\mathrm{logMAR_R},\mathrm{logMAR_L})^{\top}$, where $\mathrm{AOA_R}$ and $\mathrm{AOA_L}$ denote the amplitudes of accommodation of the right and left eyes, and $\mathrm{logMAR_R}$ and $\mathrm{logMAR_L}$ denote the corresponding unaided visual acuities transformed from Snellen measurements to the standard logMAR scale. The response vector is $\bY=(\mathrm{SE_R},\mathrm{SE_L},\mathrm{IOP})^{\top}$, where $\mathrm{SE_R}$ and $\mathrm{SE_L}$ are the spherical equivalents of the right and left eyes, defined as $\mathrm{sphere}+0.5\times\mathrm{cylinder}$, and $\mathrm{IOP}$ is the mean intraocular pressure across the two eyes. We remove observations with missing or invalid selected measurements, including nonpositive amplitudes of accommodation and invalid visual-acuity measurements, resulting in 2,430 complete observations for the analysis. All retained covariates and responses are then standardized to have zero sample mean and unit sample variance.
	
	We seek to assess whether the covariates $\bX$ are associated with the multivariate ophthalmic response $\bY$ and, if so, to characterize both the association and the conditional distribution of $\bY$ given $\bX$. To this end, we apply the proposed independence test, together with the competing tests considered in Section~\ref{sec-simulation-independence}, to the eye-screening data. All four tests yield a permutation $p$-value of 0.0099 and therefore reject the null hypothesis of independence at the 5\% significance level, providing consistent evidence of an association between $\bX$ and $\bY$.
	
\begin{figure}[htbp]
		\centering
		\includegraphics[width=0.8\textwidth]{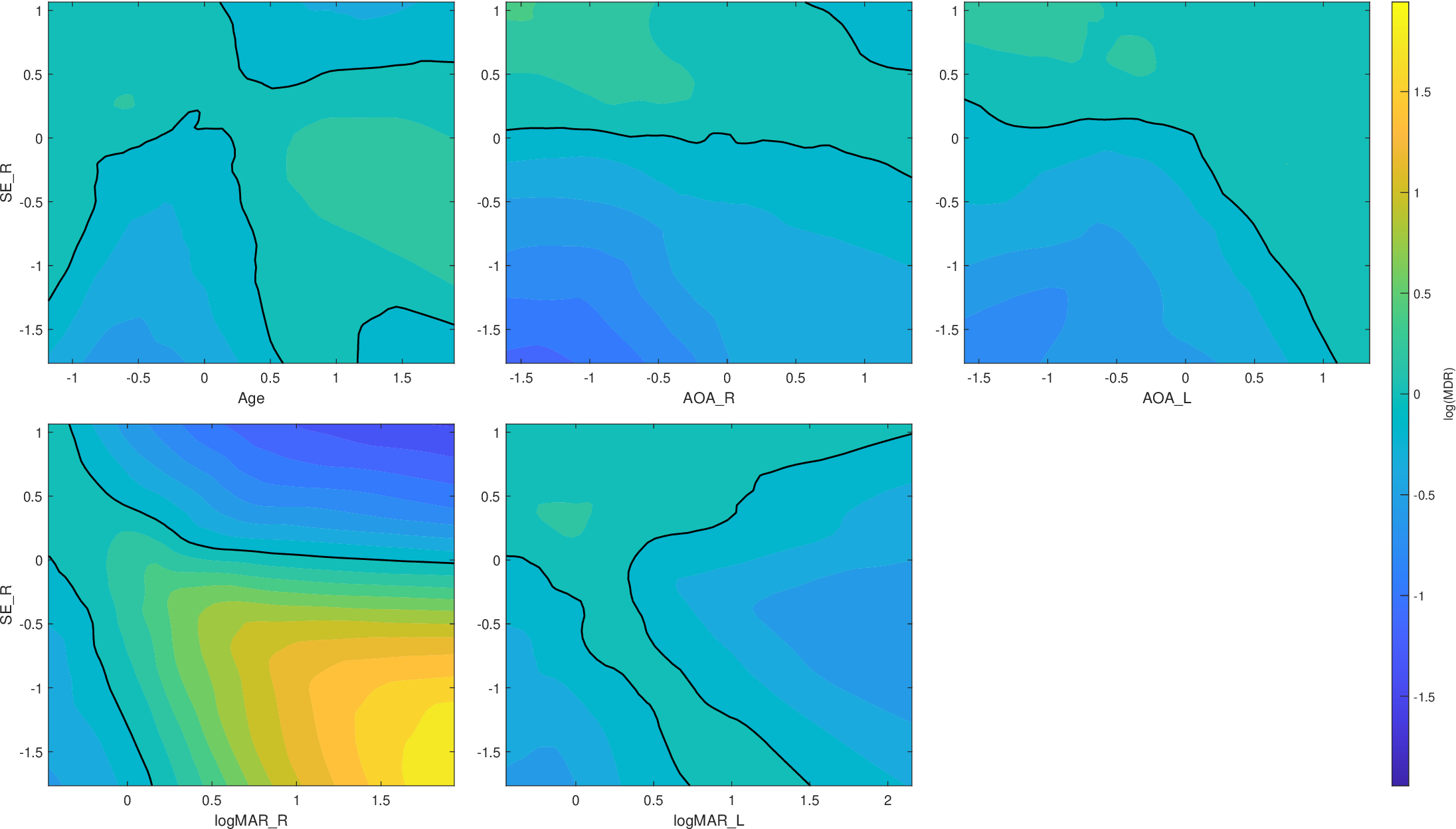}
		\caption{Estimated log-MDR slices relating $\mathrm{SE_R}$ to each covariate in the Eye Screening Data under the proposed method, with all non-displayed variables fixed at their sample medians. Black curves indicate the zero contours of the log-MDR.}
		\label{fig:mdr-ser}
	\end{figure}

	To further investigate the association, we apply the MDR and MI estimators considered in Section~\ref{sec-simulation-conditional-mean}. The estimated MI between $\bX$ and $\bY$ is 0.185 for the proposed method, compared with 0.139 for Logistic, 0.225 for Least-squares, and 0.440 for the Plug-in estimator. The positive estimates are consistent with the rejection of independence, although the estimated magnitude of dependence varies across methods. Figure~\ref{fig:mdr-ser} presents two-dimensional slices of the estimated log-MDR for $\mathrm{SE_R}$ obtained by the proposed method, with all non-displayed components fixed at their sample medians. The black curves indicate the zero level of the log-MDR and thus separate regions where the local joint density lies above or below the independence benchmark. The estimated dependence patterns vary substantially across covariates.
	
	The most pronounced pattern emerges for the $\mathrm{logMAR_R}$-$\mathrm{SE_R}$ pair. Recall that a larger $\mathrm{logMAR_R}$ indicates poorer unaided visual acuity, whereas a more negative $\mathrm{SE_R}$ corresponds to a more myopic refractive state. The estimated log-MDR is distinctly positive in regions combining relatively large $\mathrm{logMAR_R}$ with negative $\mathrm{SE_R}$, and becomes negative for combinations of large $\mathrm{logMAR_R}$ and higher $\mathrm{SE_R}$. This suggests a stronger local association between poorer unaided visual acuity and a more myopic refractive state relative to the independence benchmark. In comparison, the dependence patterns of $\mathrm{SE_R}$ with age, accommodative amplitudes, and left-eye logMAR are generally more moderate, although their positive and negative regions also vary across the covariate-response space.
	
	\begin{table}[!ht]
		\centering
		\caption{Prediction errors for conditional features based on the Eye Screening Data. Mean squared error is reported for conditional means and conditional product means, while check loss is reported for conditional medians.}
		\label{tab:eye-conditional-features}
		\setlength{\tabcolsep}{3.5pt}
		\renewcommand{\arraystretch}{1.05}
		\resizebox{\textwidth}{!}{
			\begin{tabular}{lccccccccc}
				\toprule
				& \multicolumn{3}{c}{Conditional mean}
				& \multicolumn{3}{c}{Conditional median}
				& \multicolumn{3}{c}{Conditional product mean} \\
				\cmidrule(lr){2-4}\cmidrule(lr){5-7}\cmidrule(lr){8-10}
				Method
				& $\mathrm{SE_R}$ & $\mathrm{SE_L}$ & $\mathrm{IOP}$
				& $\mathrm{SE_R}$ & $\mathrm{SE_L}$ & $\mathrm{IOP}$
				& $\mathrm{SE_R}\mathrm{SE_L}$
				& $\mathrm{SE_R}\mathrm{IOP}$
				& $\mathrm{SE_L}\mathrm{IOP}$ \\
				\midrule
				Proposed
				& 0.8411 & \textbf{0.8355} & 1.0769
				& \textbf{0.2895} & \textbf{0.3058} & 0.4101
				& \textbf{7.3540} & \textbf{0.9350} & \textbf{1.2358} \\
				
				Logistic
				& \textbf{0.8375} & 0.8529 & \textbf{1.0678}
				& 0.2906 & 0.3067 & 0.4072
				& -- & -- & -- \\
				
				Least-squares
				& 0.8542 & 1.0664 & 1.0873
				& 0.2929 & 0.3317 & 0.4103
				& -- & -- & -- \\
				
				LinCDE
				& 0.8982 & 0.8664 & 1.0700
				& 0.3054 & 0.3101 & \textbf{0.4059}
				& -- & -- & -- \\
				
				NNKCDE
				& 0.9662 & 0.9589 & 1.0688
				& 0.3111 & 0.3220 & 0.4067
				& 10.1952 & 0.9352 & 1.2396 \\
				
				DRF
				& 0.8618 & 0.8692 & 1.0708
				& 0.2957 & 0.3122 & 0.4108
				& 8.5362 & 0.9388 & 1.2366 \\
				\bottomrule
		\end{tabular}}
	\end{table}
	
	In addition, the conditional distribution of $\bY$ given $\bX$ fully characterizes the association between them. To evaluate the predictive accuracy of the conditional CDF estimation methods considered in Section~\ref{sec-simulation-conditional-mean}, we randomly split the data into a 70\% training sample and a 30\% test sample. The training sample is used to fit the conditional CDF for each method, and the resulting conditional features are assessed on the test sample, with the same split employed for all methods. Table~\ref{tab:eye-conditional-features} compares the prediction errors of the estimated conditional features: mean squared error (MSE) is used for conditional means and conditional product means, and the quantile check loss for conditional medians.
	
	The proposed method performs competitively for the marginal conditional means and medians and attains the smallest prediction errors for all three conditional product means. The improvement is particularly clear for $\e(\mathrm{SE_R}\,\mathrm{SE_L}\mid\bX)$, where the proposed method achieves an MSE of 7.354, compared with 8.536 for DRF and 10.195 for NNKCDE. It is also worth noting that the proposed method, NNKCDE, and DRF estimate the joint conditional distribution of $\bY$ and thus obtain all marginal conditional means and medians from a single fitted model, whereas Logistic, Least-squares, and LinCDE require separate conditional-density fits for each response component. These results suggest that the proposed method retains competitive accuracy for marginal conditional features while offering a clear advantage in estimating features that involve the joint conditional distribution.

	\section{Discussion}
	\label{sec-discussion}
	
	We have proposed an NETM for the conditional density of $\bY$ given $\bX$, in which the marginal baseline of $\bY$ is left unspecified and the dependence is carried by a nonparametric interaction function $\phi(\bx,\by)$. The model extends the GLM in two directions at once: the baseline is profiled out by empirical likelihood, as in the proportional likelihood ratio model of \cite{luo2012}, and the natural parameter is replaced by a nonparametric interaction estimated over a sieve. The resulting profile likelihood framework yields, from a single fit, estimators of the MDR, the MI, conditional means, quantiles, and distribution functions, together with a permutation test of independence that is valid in finite samples. Under FNN sieves and a hierarchical composition assumption on $\phi_0$, all of these estimators attain rates governed by the intrinsic smoothness--dimension ratio $\gamma$ in \eqref{intrinsic-dim-ratio} rather than by the ambient dimension. Compared with the Bregman-divergence approach of \cite{Han2025}, which requires a separate estimate of the marginal density of $\bY$ to convert the mutual density ratio into a conditional density, our likelihood approach delivers a properly normalized conditional distribution directly. Our simulations indicate that this translates into smaller estimation error in most settings considered.
	
	Several directions merit further study. First, our error bounds are stated in $L_2(P_X P_Y)$ and are of the same order as the sieve rate; establishing asymptotic normality and valid confidence intervals for smooth functionals such as the mutual information or a conditional mean at a fixed point would require a second-order expansion of the profile likelihood and a debiasing step, in the spirit of \cite{chernozhukov2018}. Second, the permutation test is exactly valid, but its power depends on the choice of sieve; a data-driven selection of the network architecture, for instance by sample splitting, would be of practical interest. Third, the representation may extend to censored or truncated responses, where the empirical likelihood for the baseline can be replaced by its Kaplan--Meier counterpart, and to the case where $\bY$ is discrete or mixed. Finally, the U-process tail bound developed here may be of independent interest for other statistical problems whose profile likelihood takes the form of a U-statistic, such as pairwise likelihood \citep{VarinReidFirth2011} and rank-based methods \citep{Lehmann2006}.

\bibliographystyle{imsart-number} 
\bibliography{entropy-ref}       

\end{document}